\documentclass[twocolumn]{aa}
\usepackage{graphicx}
\usepackage{txfonts}
\begin{document}
   \title{X-ray spectra of {\it XMM-Newton} serendipitous medium flux
   sources} \titlerunning{X-ray spectra of medium sources}

   \subtitle{}
   \author{
	 S. Mateos	\inst{1,5} 
	\and
	 X. Barcons	\inst{1} 
	\and
	 F. J. Carrera	\inst{1}
	\and
	 M. T. Ceballos	\inst{1}
	\and
	A. Caccianiga	\inst{2}
 	\and
         G. Lamer	\inst{3} 
	\and
	T. Maccacaro	\inst{2}
	\and
	 M.J. Page	\inst{4}
	\and
         A. Schwope	\inst{3} 
	\and
         M. G. Watson 	\inst{5}
	}

           \authorrunning{S. Mateos et al.}

   \offprints{S. Mateos, \email{sm279@star.le.ac.uk}}
   \institute{Instituto de F\'\i sica de Cantabria (CSIC-UC), 39005 Santander, Spain 
	\and Osservatorio Astronomico di Brera, via Brera 28, 20121 Milano, Italy
	\and Astrophysikalisches Institut Potsdam, An der Sternwarte 16, D-14482 Potsdam, Germany
	\and Mullard Space Science Laboratory, UCL, Holmbury St Mary, Dorking, Surrey RH5 6NT, UK 
	\and Department of Physics and Astronomy, University of Leicester, LE1 7RH, UK}

 \date{15 December 2004}

   \abstract{We report on the results of a detailed analysis of
   the X-ray spectral properties of a large sample of sources detected
   serendipitously with the {\it XMM-Newton} observatory in 25 selected
   fields, for which optical identification is in progress.  The
   survey covers a total solid angle of $\sim$ 3.5 $\mathrm{deg}^{2}$
   and contains 1137 sources with $\sim 10^{-15}<{\it
   S}_{0.5-10}<10^{-12}\,{\rm erg\,cm}^{-2}\,{\rm s}^{-1}$ with good 
   enough spectral quality as to perform a detailed X-ray spectral analysis of 
each individual object. We find
   evidence for hardening of the average X-ray spectra of the sources towards
   fainter fluxes, and we interpret this as indicating a higher degree of 
   photoelectric
   absorption amongst the fainter population.  Absorption is detected at 95\% 
   confidence in 20\% of
   the sources, but it could certainly be 
   present in many other sources below our detection capabilities.  For Broad 
   Line AGNs (BLAGNs), 
   we detect absorption in $\sim 10\%$ of the sources 
   with column densities in the range $10^{21}-10^{22}\, {\rm cm}^{-2}$. The fraction of 
   absorbed Narrow Emission Line galaxies (NELGs, most with intrinsic 
   X-ray luminosities $> 10^{43}\,{\rm erg\,s^{-1}}$, and 
   therefore classified as type 2 AGNs) is significantly higher (40\%), with a hint
   of moderately higher columns. After correcting for absorption, we
   do not find evidence for a redshift evolution of the underlying
   power law index of BLAGNs, which stays roughly constant at
   $\Gamma\sim 1.9$, with intrinsic dispersion of $0.4$.  A small
   fraction ($\sim 7\%$) of BLAGNs and NELGs require the presence of 
   a soft excess, that we model as a
   black body with temperature ranging from 0.1 to 0.3 keV. Comparing
   our results on absorption to popular X-ray background synthesis
   models, we find absorption in only $\sim 40\%$ of the sources
   expected. This is due to a deficiency of heavily absorbed sources
   (with ${\rm N_H}\sim 10^{22}-10^{24}\, {\rm cm}^{-2}$) in our sample in
   comparison with the models. We therefore conclude that the
   synthesis models require some revision in their specific
   parameters.

    } \maketitle
%

\section{Introduction}
\label{Introduction}

Extensive studies with different satellites have been performed during
the last decades to unveil the origin of the cosmic X-ray background
radiation (XRB).  In the 0.5-2 keV band, the {\it ROSAT} deep
surveys in the Lockman Hole (Hasinger et al. \cite{Hasinger1998})
resolved $\sim70-80\%$ of the XRB into discrete sources down to a flux
limit of $\sim10^{-15}\,{\rm erg\,cm}^{-2}\,{\rm s}^{-1}$.  The deepest
{\it Chandra} observations have now resolved $\sim 90\%$ of the soft
XRB reaching a flux limit of $\sim2\times10^{-16}\,{\rm
erg\,cm}^{-2}\,{\rm s}^{-1}$ (Moretti et al. \cite{Moretti2003}).

In the hard X-ray sky, important contributions to the population
studies were performed by the $ASCA$ 
(Georgantopoulos et al. \cite{Georgantopoulos1997}, Cagnoni, Della Ceca \& Maccacaro 
\cite{Cagnoni1998}, Boyle et al. \cite{Boyle1998}, Ueda et al. \cite{Ueda1998},\cite{Ueda1999}, 
Della Ceca et al. \cite{DellaCeca1999}) and $BeppoSAX$ (Giommi, Perri \& Fiore \cite{Giommi2000}) 
satellites. They
resolved $\sim30\%$ of the 2-10 keV hard XRB (Ueda et al
\cite{Ueda1998}, Fiore et al. \cite{Fiore2001}) down to fluxes of
$5\times10^{-14}\,\mathrm{erg\,cm}^{-2}\,\mathrm{s}^{-1}$. 
$Chandra$ and {\it XMM-Newton} have reached fluxes 100 times fainter,
resolving $\sim 50-70\%$ of the hard XRB (Mushotzky et al
\cite{Mushotzky2000}, Barger et al. \cite{Barger2001}, Fiore et al
\cite{Fiore2001}, Giacconi et al. \cite{Giacconi2001}, 
Hasinger et al. \cite{Hasinger2001}, Tozzi et al
\cite{Tozzi2001}, Baldi  et al. \cite{Baldi2002}, Worsley et al. \cite{Worsley2004}).
The fraction of resolved X-ray background has recently been increased to 
$\sim 90\%$ in the 2-10 keV and $\sim 94\%$ in the 0.5-2 keV bands by
Cowie et al. \cite{Cowie2002} and 
Moretti et al. \cite{Moretti2003}, respectively.

Optical follow-up of the faint X-ray sources has revealed that the
bulk of the XRB is made up of type 1 and type 2 AGNs. The results of  
optical identification programmes suggest that the
fractional contribution of type 2 AGNs is becoming increasingly important
towards fainter fluxes (e.g. Fiore et al. \cite{Fiore2003}).

Now that the discrete origin of the XRB has been confirmed all over
the X-ray energy band (0.1-10 keV), efforts are being concentrated on
the study of the source populations that make up the XRB. The nature
and cosmic evolution of these sources is still controversial (Barger 
et al. \cite{Barger2002}, Mainieri et al. \cite{Mainieri2002}, 
Fiore et al. \cite{Fiore2003}, 
Ueda et al. \cite{Ueda2003}).

At the faintest X-ray fluxes an important
contribution to the XRB emission is expected to come from type 2 AGN
(Setti\& Woltjer \cite{Setti1989}, Gilli et al. \cite {Gilli2001}, 
Comastri et al. \cite {Comastri1995}).  This has been confirmed 
observationally (Mainieri et al. \cite{Mainieri2002}).  
The distribution of intrinsic absorption among the
sources has only been measured observationally for the brightest
sources at low redshifts, but is still unknown at high redshifts.
Recent {\it Chandra} and {\it XMM-Newton} studies have revealed that
the widely accepted correlation between the optical obscuration and X-ray
absorption (type 1 AGN being unabsorbed and type 2 AGN being
absorbed) is not always true.  Several cases of type 1 AGN suffering
from significant X-ray absorption have been found 
(Mittaz et al. \cite{Mittaz1999}, Fiore et al. \cite{Fiore2001},
Page et al. \cite{Page2001}, 
Schartel et al. \cite{schartel}, Tozzi et al. \cite{Tozzi2001}, 
Mainieri et al. \cite{Mainieri2002}, Brusa et al. \cite{Brusa2003}, 
Page et al. \cite{Page2003}, Carrera et al. \cite{Carrera2004}, Perola et al. \cite{Perola2004}), 
as well as Seyfert 2 galaxies unabsorbed in X-rays
(Pappa et al. \cite{Pappa2001}, Panessa et al. \cite{Panessa2002},
Barcons et al. \cite{Barcons2003}).

The motivation for this work is to exploit 
the large collecting area and wide field of view (FOV)
of {\it XMM-Newton} to perform a detailed spectral analysis of X-ray selected sources down to
medium-faint fluxes.  This study will allow us to put strong
observational constraints on the average X-ray spectral properties of
the source population that dominates the X-ray emission at the flux
level of our survey, typically $\sim 10^{-14}\,\mathrm{erg\,cm}^{-2}\,
\mathrm{s^{-1}}$ in the 0.5-10 keV energy band.  Therefore we have conducted a
large (total solid angle of $\sim 3.5\,\mathrm{deg}^{2}$) and medium
to deep (0.5-10 keV flux in the range
$\sim10^{-15}-10^{-12}\,\mathrm{erg\,cm}^{-2}\,\mathrm{s^{-1}}$) survey
of sources detected with the {\it XMM-Newton} observatory in the
0.2-12 keV energy band.  We take advantage of an ongoing
identification effort of sources in the fields under study, which
enable us to obtain spectral properties of representative samples of
identified sources. This work complements an earlier study of the
X-ray spectral properties of a large sample of sources at similar
fluxes, but where the count rates detected in the XMM-Newton
standard energy bands (0.2-0.5, 0.5-2, 2-4.5, 4.5-7.5, 7.5-12
keV) were used to analyse the broad band X-ray spectral shape of the sources
(Mateos et al
\cite{Mateos2003}).

The paper is organised as follows: Sect.~\ref{X-ray observations} describes the X-ray data
and the current status of the identifications; Sect.~\ref{products} describes the
extraction of the X-ray spectral products; Sect.~\ref{spectral_analysis} describes the 
results of the spectral fitting; Sect.~\ref{sp_fitting} discusses the average
spectral properties of our sources; the X-ray spectral properties of 
sources identified as ALGs are described in Sect.~\ref{ALGs}; 
Sect.~\ref{synthesis} compares our observational results with specific XRB synthesis models; 
the results of our analysis are summarised in Sect.~\ref{conclusions}. Throughout this paper we have adopted the
currently favoured concordance cosmology parameters ${\rm
H_0=70\,km\,s}^{-1}\,{\rm Mpc}^{-1}$, $\Omega_m=0.3$ and
$\Omega_{\Lambda}=0.7$. 
The luminosity values that we use in the paper are the intrinsic 
luminosities of the objects corrected for absorption.
All one-parameter uncertainties are computed with a delta
chi-squared of 2.706, equivalent to 90\% confidence region for a
single parameter.


\section{Observations}
\label{X-ray observations}

   \begin{table*}
   \begin{minipage}{100mm} \caption[]{Observational details of the
   {\it XMM-Newton} fields}
     $$
          \begin{array}{l c c c c c c c c c c c} \hline
          \noalign{\smallskip}
           {\rm Field} & {\rm R.A.} & {\rm Dec} & {\rm b\,^{\mathrm{a}}} & {\rm
           N_H^{Gal}} & {\rm Rev./Obs.\,\,id} & & {\rm
           Filter\,^{\mathrm{b}}} & & & {\rm GTI\,^{\mathrm{c}}(ksec)}
           \\ & (J2000) & (J2000) & {\rm (deg)} & {\rm (10^{20}\, cm^{-2})} & & {\rm M1}
           & {\rm M2} & {\rm pn} & {\rm M1} & {\rm M2} & {\rm pn}\\
           \noalign{\smallskip} \hline \hline \noalign{\smallskip}
           {\rm A2690} & 00:00:30.3 &-25:07:30.0 & -78.90 & 1.84 &
           0088/0125310101 & {\rm M} & {\rm M} & {\rm M} &19 & 19 &
           21\\ {\rm CL0016+16} & 00:18:33.0 &+16:26:18.0 & -45.54 &
           4.07 &0194/0111000101 & {\rm M} & {\rm M} & {\rm M} & 31 &
           31 & 29 \\ & & & & & 0194/0111000201 & {\rm M} &{\rm M} &
           {\rm M} & 5 & 5 & 3\\ {\rm G133-69\,\,pos\,\,2} &
           01:04:00.0 &-06:42:00.0 & -69.35 & 5.19 & 0104/0112650501 &
           {\rm Th}& {\rm M} & {\rm Th} & 23 & 23 & 18 \\ {\rm
           G133-69\,\,pos\,\,1} & 01:04:24.0 &-06:24:00.0 & -68.68 &
           5.20 & 0188/0112650401 & {\rm Th} & {\rm M} & {\rm Th} & 23
           & 23 & 20 \\ {\rm SDS-1b} & 02:18:00.0 &-05:00:00.0 &
           -59.75 & 2.47 & 0118/0112370101 & {\rm Th} & {\rm Th} &
           {\rm Th} & 47 & 48 & 40 \\ & & & & & 0119/0112371001 & {\rm
           Th} & {\rm Th} & {\rm Th} & 51 & 51 & 43 \\ {\rm SDS-3} &
           02:18:48.0 &-04:39:13.0 & -59.35 & 2.54 & 0121/0112370401 &
           {\rm Th} & {\rm Th} & {\rm Th} & 21 & 21 & 15 \\ & & & & &
           0121/0112371501 & {\rm Th} & {\rm Th} & {\rm Th} & 7 & 7 &
           4 \\ {\rm SDS-2} & 02:19:36.0 &-05:00:00.0 & -58.91 & 2.54
           & 0120/0112370301 & {\rm Th} & {\rm Th} & {\rm Th} & 50 &
           50 & 40 \\ {\rm Mkn3} & 06:15:36.3 &+71:02:15.0 & 22.72 &
           8.82 & 0158/0111220201 & {\rm M} & {\rm M} & {\rm M} & 54 &
           54 & 44 \\ {\rm MS0737} & 07:44:04.5 &+74:33:49.5 & 29.57 &
           3.51 & 0063/0123100101 & {\rm Th} & {\rm Th} & {\rm Th} &
           39 & 39 & 20 \\ & & & & & 0063/0123100201 & {\rm Th} & {\rm
           Th} & {\rm Th} & 19 & 19 & 20 \\ {\rm CL0939+472} &
           09:43:00.0 &+46:59:30.0 & 48.88 & 1.24 & 0167/0106460101 &
           {\rm Th} & {\rm Th} & {\rm Th} & 49 & 49 & 43 \\ {\rm
           S5\,\,0836+716} & 08:41:24.0 &+70:53:41.0 & 34.43 & 2.98 &
           0246/0112620101 & {\rm M} & {\rm M} & {\rm M} & 4 & 4 & 25
           \\ {\rm B2\,\,1028+31} & 10:30:59.1 &+31:02:56.0 & 59.79 &
           1.94 & 0182/0102040301 & {\rm Th} & {\rm Tck} & {\rm Th} &
           26 & 26 & 23 \\ {\rm B2\,\,1128+31} & 11:31:09.4
           &+31:14:07.0 &72.03 & 2.00 & 0175/0102040201 & {\rm Th} &
           {\rm Tck} & {\rm Th} & 19 & 23 & 13 \\ {\rm Mkn205} &
           12:21:44.0 &+75:18:37.0 & 41.67 & 3.02 & 0075/0124110101 &
           {\rm M} & {\rm M} & {\rm M} & 50 & 50 & 37 \\ {\rm
           MS1229.2+6430} & 12:31:32.0 &+64:14:21.0 & 53.05 & 1.98 &
           0082/0124900101 & {\rm Th} & {\rm Th} & {\rm Th} & 34 & 34
           & 29 \\ {\rm HD\,\,117555} & 13:30:47.0 &+24:13:59.0 &
           80.67 & 1.16 & 0199/0100240101 & {\rm M} & {\rm M} & {\rm
           M} & 29 & 29 & 25 \\ & & & & & 0205/0100240201 &{\rm M} &
           {\rm M} &{\rm M} & 36 & 36 & 33 \\ {\rm UZ\,\,Lib} &
           15:32:23.0 &-08:32:05.0 & 36.57 & 8.97 & 0210/0100240801 &
           {\rm M} & {\rm M} & {\rm M} & 23 & 23 & 23 \\ {\rm
           PKS\,\,2126-158} & 21:29:12.2 &-15:38:41.0 & -42.39 & 5.00
           & 0255/0103060101 & {\rm M} & {\rm M} & {\rm M} & 22 & 22 &
           16 \\ {\rm PKS\,\,2135-147} & 21:37:45.2 &-14:32:55.4 &
           -43.85& 4.70 & 0254/0092850201 & {\rm M} & {\rm M} & {\rm
           M} & 15 & 16 & 28 \\ {\rm MS2137-23} & 21:40:15.2
           &-23:39:41.0 & -47.50 & 3.50 & 0254/0008830101 & {\rm Th} &
           {\rm Th} & {\rm Th} & 13 & 14 & 10 \\ {\rm PB5062} &
           22:05:09.8 &-01:55:18.0 & -43.28 & 6.17 & 0267/0012440301 &
           {\rm Th} & {\rm Th} & {\rm Th} & 31 & 31 & 28 \\ {\rm
           LBQS\,\,2212-1759} & 22:15:31.6 &-17:44:05.7 &-52.92 & 2.39
           & 0355/0106660401 & {\rm Th} & {\rm Th} & - & 33 & 33 & -
           \\ & & & & & 0355/0106660501 & {\rm Th} & {\rm Th} & {\rm
           Th} & 8 & 5 & 8 \\ & & & & & 0356/0106660601 & {\rm Th} &
           {\rm Th} & {\rm Th} &103 & 103 & 91 \\ {\rm PHL\,\,5200} &
           22:28:30.4 &-05:18:55.0 &-49.97 & 5.26 & 0269/0100440101 &
           {\rm Tck} & {\rm Tck} & {\rm Tck} & 45 & 45 & 43 \\ {\rm
           IRAS22491-18} & 22:51:49.4 &-17:52:23.2 & -61.42 & 2.71 &
           0267/0081340901 & {\rm M} & {\rm M} & {\rm M} & 22 & 22 &
           20 \\ {\rm EQ\,\,Peg} & 23:31:50.0 &+19:56:17.0 & -39.14 &
           4.25 & 0107/0112880301 & {\rm Tck} & {\rm Tck} & {\rm Tck}
           & 14 & 14 & 12 \\ \noalign{\smallskip} \hline \end{array}
           $$
         \label{tab0}
\begin{list}{}{}
\item[$^{\mathrm{a}}$] Galactic latitude

\item[$^{\mathrm{b}}$] Blocking filters: Th: Thin at 40nm A1; M: Medium at 80nm A1; 
		       Tck: Thick at 200nm A1
\item[$^{\mathrm{c}}$] Good time intervals after removal of flares
\end{list}
\end{minipage}
\end{table*}

In this study we use 25 observations selected from the public {\it
XMM-Newton} data archive. They cover a total solid angle of
$\sim$3.5 deg$^2$.  To obtain a clean extragalactic sample of
objects and to maximise the survey efficiency, the X-ray observations
were selected according to the following criteria: sky positions at
high galactic latitudes ($|b|>20\degr$); observations for which 
we had data from the European Photon Imaging Camera (EPIC)-pn detector in
FULL-FRAME MODE (i.e. the EPIC-pn full FOV covered). We avoided fields
containing bright point or extended 
targets where the wings of the
target can reduce substantially the area of the survey covered.  The current
analysis has been performed within the framework of the tasks of the
{\it XMM-Newton} Survey Science Centre\footnote{See {\tt
http://xmmssc-www.star.le.ac.uk}}.

The X-ray fields are listed in Table~\ref{tab0} together with their
observational details.

\subsection{X-ray Data reduction} 
\label{Data reduction} 

All the observations were processed through the {\it XMM-Newton}
pipeline processing system that uses a suite of tasks created
specifically for the analysis of {\it XMM-Newton} data, namely the
Science Analysis System (Gabriel et al. \cite{Gabriel2004}).
 The current version at the time of the analysis 
being v5.3.3. 
The pipeline process provides
calibrated event files, images and exposure maps in five standard
X-ray energy bands (0.2-0.5, 0.5-2, 2-4.5, 4.5-7.5 and 7.5-12
keV) covering the energy range where the data is best calibrated.  The
X-ray images are created with a pixel size of 4\arcsec x 4\arcsec and
cover a FOV of $\sim$ 30$\arcmin$ diameter.  The exposure maps give
information on the spatial efficiency of each camera including the
energy dependent mirror vignetting.  The pipeline products also provide source lists
for each observation and for each EPIC camera (pn, MOS1 and MOS2).

The source detection is run independently for each of the EPIC
detectors, simultaneously on the five energy bands defined above. The
area of the images where the sources are searched for is defined by
the SAS task {\tt emask}, that creates a map where CCD gaps and bad
pixels/columns are excluded.  The source detection process starts with
the task {\tt eboxdetect} (local mode) that performs a sliding box cell
detection of sources with a minimum likelihood of detection of 8.
This provides an input list of source positions masked out by the SAS
task {\tt esplinemap} to build background maps for each detector and energy
band.  {\tt Eboxdetect} (map mode) is run again, but this time the
background is taken from the maps created by {\tt esplinemap}. This improves
significantly the detection sensitivity.  Finally the task {\tt emldetect}
performs maximum likelihood Point Spread Function (PSF)\footnote{{\it
XMM-Newton} mirror modules have a PSF with FWHM of $\sim$6$\arcsec$}
fits to the distribution of counts of the sources detected by
{\tt eboxdetect}, obtaining for each object the total (0.2-12 keV) count rate
and the likelihood of detections (total and on each energy band). For a 
description of the application of Maximum Likelihood analysis to detection of sources see
Cruddace et al. \cite{Cruddace1987}. The sources with a total detection likelihood higher than 10 are included
in the final list.

\subsection{X-ray source list}
\label{X-ray source list}

The geometrical shadowing of half of the sky X-ray emission received
by the MOS1 and MOS2 detectors (deviated to RGS gratings) makes
the EPIC-pn camera a factor of two more sensitive
than the MOS cameras overall. Therefore, to optimise the sensitivity of the survey, 
we used the pipeline EPIC-pn source lists. In the cases where we had more than 
one observation for a field we used the source list of the deepest exposure.

We defined circular regions (with radius between 16 to 160 arcsec), to
exclude from the images the areas contaminated by the emission of the
targets. For the objects detected near
CCD gaps the uncertainties in their coordinates and flux 
may be larger than the statistical uncertainties as given by the 
pipeline products. In order to get rid of these sources with somewhat 
uncertain source parameters, we have excluded zones near 
CCD gaps, widening them in size up to the
radius that contains $80\%$ of the PSF on each detector point. We then
used these masks to remove from the source list all the objects
detected close to CCD gaps. 
For each field and observation we have examined the images 
searching for spurious sources, like
detections in hot pixels or bright segments not removed by the
pipeline process. We have checked carefully detections close to
the tails of bright and/or extended objects, that in most cases turned
out to be spurious.  The visual screening process removed $\sim10\%$
of the objects from the original list.  
The total number of sources detected within these specifications was 2145.

\subsection{Source detection sensitivity}
\label{Catalogue biases}
The variety of data that we have used in our study, 
make it difficult to calculate the source detection sensitivity as a function of the 
X-ray flux, because in principle it depends on the exposure times of the fields and their Galactic absorbing 
columns, $\rm {N_H}$ (Zamorani et al. \cite{Zamorani1988}).

The {\it XMM-Newton} EPIC-pn detector is sensitive to X-ray photons at energies ranging from $\sim$ 0.2 to $\sim$ 12 keV. 
The sensitivity is maximum between 0.5 and 4.5 keV, where it is almost independent of energy, i.e., objects with different spectral shapes are 
detected with similar efficiency. Outside this energy interval, the 
sensitivity is a strong function of energy.

Because the band of detection of photons is wide, we do not expect the use of fields with 
different $\rm {N_H^{Gal}}$ (ranging from 1 to $9\times 10^{20}\, {\rm
cm^{-2}}$) to introduce any significant bias (eg. detections in the fields
with the higher values of Galaxy absorption biased towards hard
\footnote{We use the term ``soft spectra''  when the measured spectral 
photon index, $\Gamma$, is found to be $\sim$2, the typical value for 
unabsorbed AGNs. We consider a spectra to be hard if $\Gamma$ $\le$1.5} 
spectrum sources) in the broad band spectral properties of our objects at the
flux limits of our survey. 

However, we can see in Table~\ref{tab0}, that the exposure times 
vary significantly from field to field. 
At 0.5-10 keV fluxes below $\sim {\rm 10^{-14}\,erg\,cm^{-1}\,s^{-1}}$ , 
where only a few fields contribute to the source list, 
we expect our survey to be biased against objects with spectra peaking 
outside the 0.5-4.5 keV energy range, for example highly obscured AGNs.

To build our source lists, the {\it XMM-Newton} source detection algorithm was run simultaneously in the five standard
{\it XMM-Newton} energy bands. Therefore we have analysed sources detected in at least one of these energy bands.
This might be important, because the sensitivity varies significantly from band to band, and hence, 
different source populations may contribute at different flux levels.

In order to calculate the source detection efficiency function, 
we have carried out simulations. A detailed description of how we obtained this 
selection function is given in appendix~\ref{apendix_A}. There is indeed a mild 
dependence of the selection function on the spectral slope $\Gamma$ at any 
given flux, and we will correct for this effect.

\begin{figure}
  \centering \hbox{
  \includegraphics[angle=90,width=8.5cm]{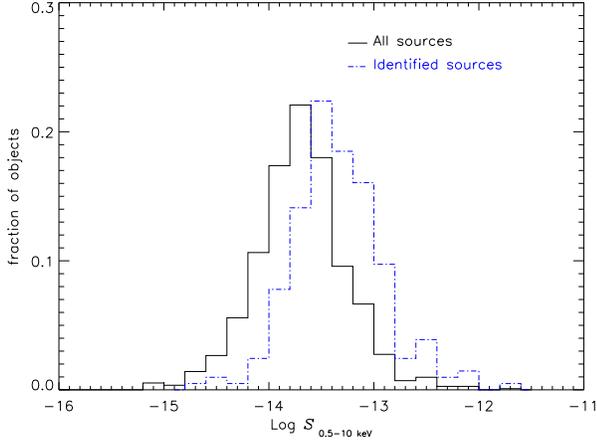}}
  \caption{$S_{0.5-10\,{\rm keV}}$ distribution for the whole sample
  (solid line) and the sub-sample of identified sources
  (dashed-dot line).}  \label{fdist}
\end{figure}

Fig.~\ref{fdist} shows the distribution of our sources 
in bins of 0.5-10 keV flux. The fluxes 
were obtained from the best fit model of each object 
(see Sect.~\ref{best_fit_model}). The selection of fields 
with a range of exposure times results in a gradual
reduction of the sky coverage as we go to fainter fluxes, and is responsible
for broadening the flux distribution below
$\sim 10^{-14}$~erg~cm$^{-2}$~s$^{-1}$.
Our sample is likely to be representative of the
dominant X-ray source population down to a flux of $\sim 10^{-14}\,{\rm
erg\,cm}^{-2}\,{\rm s}^{-1}$. For comparison we have also plotted
in Fig.~\ref{fdist} the distribution of flux for the sources with optical spectroscopic 
identifications.

The goal of the present paper is not to produce complete
source lists to build up source counts or luminosity functions (which
will be the subject of a forthcoming paper), but rather to study the
X-ray spectral properties of representative samples within the medium
flux range, where it is known that $\sim 50\%$ of the 0.5-2 keV 
(and $\sim 40\%$ of the 2-8 keV, see Bauer et al. \cite{Bauer2004})
accretion power in the Universe is produced.  For this purpose, none 
of the above
possible biases in the source list has any significant impact on our 
results, after appropriate corrections are applied.

\subsection{Optical spectroscopic identifications}
\label{Optical observations}

The selected fields are being followed-up in the optical band as part
of the {\it XMM-Newton} Survey Science Centre XID identification programme
(Watson et al. \cite{Watson2001}). The observing time
was awarded by a project named
AXIS\footnote{http://venus.ifca.unican.es/$\sim$xray/AXIS/} ("An {\it
XMM-Newton} International Survey", Barcons et al
\cite{Barcons2002}).  The optical observations consist of high quality
multicolour imaging and optical spectra of the brightest sources (the
optical observations and spectroscopic identifications 
will be described in detail in a forthcoming paper). This is an
ongoing process, whose first results were presented in Watson et al. 
\cite{Watson2001}
and Barcons et al. \cite{Barcons2002}. The full list of identified
sources will be made available via the {\it XMM-Newton} Science
Archive when completed down to a 0.5-4.5 keV flux of $2\times
10^{-14}\, {\rm erg}\, {\rm cm}^{-2}\, {\rm s}^{-1}$. 
Two identifications in the A2690 field were from Piconcelli et al. 
\cite{Piconcelli2002}.
For the time being, $\sim$ 11\% (232) of the sources in our sample have been 
identified via
optical spectroscopy. We have optical identifications within all
fields except one (MS1229), the fraction of identified sources
per field ranging from $\sim1.4\%$ to $\sim39\%$.

\begin{table}
\caption{Breakdown of the sources with optical spectroscopic identifications. 
(BLAGN: broad line AGNs; 
NELG: narrow emitting line galaxies; ALG: absorption line galaxies)}
\begin{tabular}{l c c}
\hline
{\rm Object\,\,class }& N$^{\mathrm{a}}$ & N good$^{\mathrm{b}}$\\
\hline
\hline
{\rm BLAGN} & 149 & 141\\ {\rm NELG} & 32 & 29\\ {\rm ALG} & 10 & 7\\
{\rm Stars} & 40 & 32\\ {\rm BL Lac} & 1 & 1\\ {\rm Total} & 232 & 210\\
\hline
\end{tabular}

\begin{list}{}{}
\item[$^{\mathrm{a}}$] Number of sources identified
\item[$^{\mathrm{b}}$] Number of sources identified that fulfil 
the quality thresholds applied to the spectra (see Sect.~\ref{products})
\end{list}
\label{tab1}
\end{table}

The objects were classified according to their optical emission line 
properties. Sources with spectra showing strong broad emission lines
(${\rm FWHM > 1000\,km\,\,s^{-1}}$) were
classified as Broad Line AGNs (BLAGNs). If only narrow emission lines
were seen (${\rm FWHM < 1000\,km\,\,s^{-1}}$), the sources were
classified (purely in optical spectroscopic terms) as Narrow Emission
Line Galaxies (NELGs). Objects with galaxy-like optical spectra and no
apparent emission lines were classified as absorption line galaxies
(ALGs), some of which are associated to clusters.  Table~\ref{tab1}
lists the number of identified sources, along with those which have
good quality X-ray spectra (see below).

Of all the identified sources, we have only analysed the spectra of 
BLAGNs, NELGs and ALGs.

\begin{figure}
  \centering \hbox{ 
  \includegraphics[angle=90,width=8.5cm]{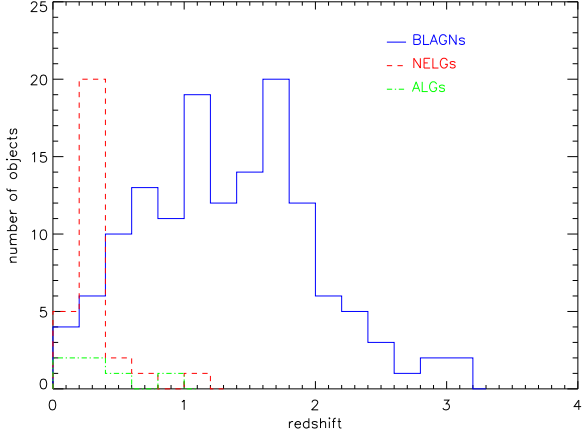}}
  \caption{ BLAGNs (solid line), NELGs (dashed line) and ALGs
  (dashed-dot line) redshift distributions.}
  \label{zdist}
\end{figure}

Fig.~\ref{zdist} shows the redshift distributions of these sources.
BLAGNs are detected up to
redshifts of $\sim3$, while at the flux limit of our optical
identifications we do not detect NELGs and ALGs at redshifts above
$\sim0.5$. Similar distributions, at similar fluxes, have been found 
previously (e.g. Green et al. \cite{Green2004} for the ChaMP survey).
In spite of the incomplete identifications, we do
expect the identified sources to constitute a "representative" sample
of the underlying population of sources at X-ray fluxes above 
$\sim 2\times 10^{-14}{\rm erg\,cm^{-2}\,s^{-1}}$, where the identified fraction of objects 
is much higher (~50\%).


\section{Extraction of X-ray spectral products}
\label{products}

We have cleaned the calibrated event files with the same filtering
expressions that were used by the pipeline to create the EPIC X-ray
images.  This filtering removes time intervals contaminated by
soft proton flares, cosmic ray tracks and spurious noise events not
created by X-rays.  For each object in the catalogue the SAS task {\tt
region} was used to define the source and background extraction
regions.  

The source region shape was defined as a circle centred on
the source coordinates and with a typical radius of $\sim14-30 \arcsec$, 
depending on the source position within the detector.
The background region
was defined as an annulus centred at the source position, with inner
radius equal to the source extraction radius and outer radius three
times the inner radius. If there was overlap with nearby sources, the 
sizes of source and/or background regions were reduced to eliminate 
the overlap.
If nearby sources were inside the background region their extraction regions 
were excluded from the background region.

Source and background spectra were then
extracted with the SAS task {\tt evselect}.  Single and double events
were included for the pn detector and single-quadruple events for the
MOS detectors in the energy band 0.2-12 keV. Further, only events
with the highest spectral quality ({\tt FLAG=0}) were included for the
EPIC-pn data.

In order to perform a proper spectral analysis, we created for each
source the redistribution matrix file (RMF) and the ancillary response
file (ARF), using the SAS tasks {\tt rmfgen} and {\tt arfgen} respectively.
To maximise the signal to noise ratio, we have combined MOS1 and
MOS2 source and background spectra and the corresponding response
matrices. Merged source and background spectra were obtained
by adding the individual spectra. Backscale values (size of the regions 
used to extract the spectra) and calibration matrices 
for the combined spectra were obtained weighting the input data  
with the exposure times.

For the fields with more than one
observation, we added all the MOS and pn data for each epoch but 
MOS and pn data were not merged because of their different responses 
\footnote{see {\it XMM-Newton} handbook at http://xmm.vilspa.esa.es}.

In order to permit the $\chi^2$ minimisation technique in the fitting
process, the raw spectra were binned such that each bin contained
$\ge10$ counts (source plus background).  We then imposed two
quality thresholds on the grouped spectra selected for 
analysis: number of bins (MOS+pn) $\ge5$, and number of 
background subtracted counts (MOS+pn) $\ge50$.

This filtering process resulted in a final sample of 1137 
sources having X-ray spectra of sufficient quality to analyse them 
individually (see Table~\ref{tab1} for the
distribution of identified sources).

\begin{figure} \centering \hbox{ 
  \includegraphics[angle=90,width=8.5cm]{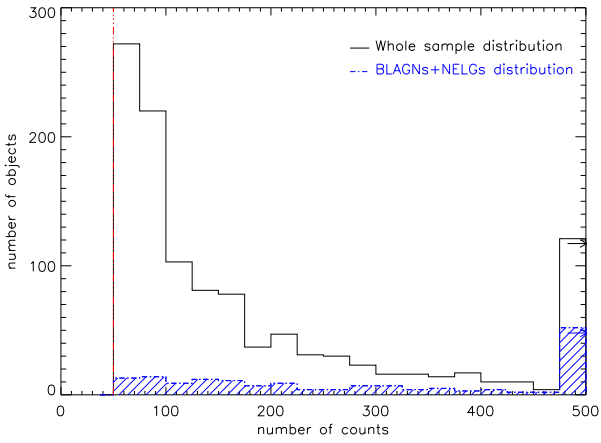}}
  \caption{Distributions of background subtracted counts (MOS+pn)
  for the objects in our sample (solid histogram) and for
  the objects classified as BLAGNs or NELGs (shaded dot-dashed histogram).
  The vertical line indicates one of the quality thresholds applied to
  the spectra.  For clarity, the last bin in the plots includes the
  spectra with more than 500 counts.}
\label{nbins}
\end{figure}

\begin{figure}
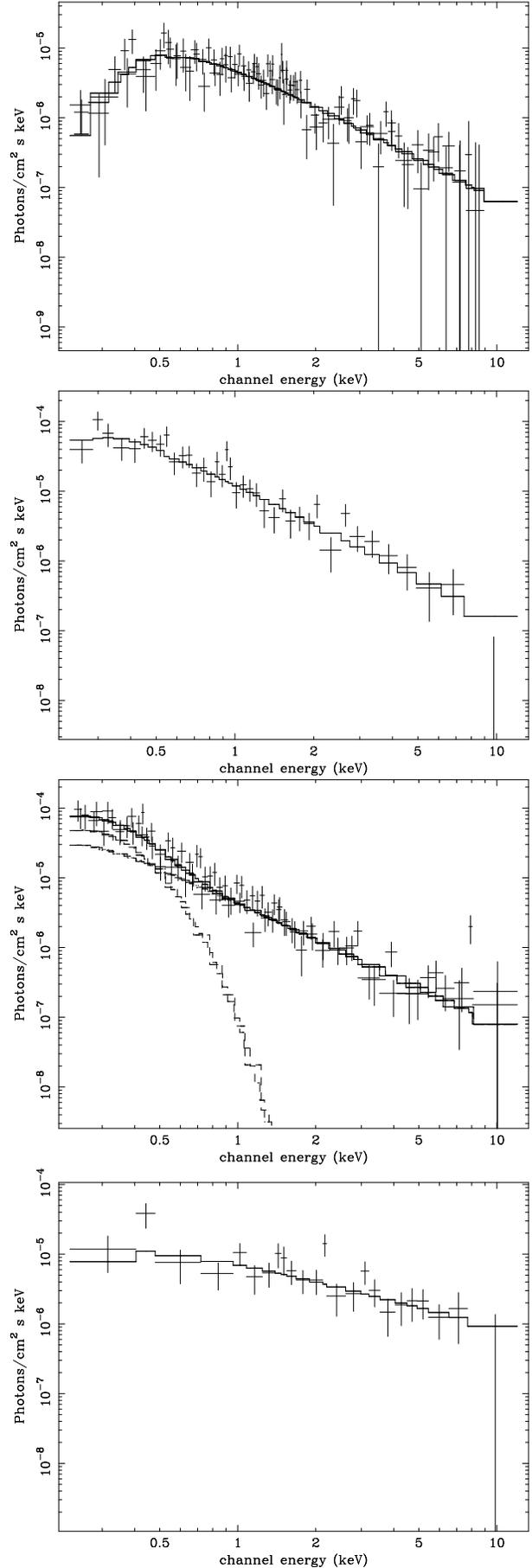

  \centering
  \hbox{
   \includegraphics[angle=-90,width=8.cm]{fig4.ps}}
   \hbox{	
   \includegraphics[angle=-90,width=8.cm]{fig5.ps}}
   \hbox{	
   \includegraphics[angle=-90,width=8.cm]{fig6.ps}}
   \hbox{	
   \includegraphics[angle=-90,width=8.cm]{fig7.ps}}
   \caption{From top to bottom unfolded spectra of a) {\bf XMMU J221454.9-173949}, absorbed BLAGN; 
	b) {\bf XMMU J084132.3+704757}, unabsorbed  NELG ($\rm{L_{2-10\,keV}\sim6\,10^{42}\,erg\,s^{-1}}$);
	 c) {\bf XMMU J33106.2+241325}, BLAGN with soft excess component (see Sect.~\ref{soft_excess} for details);
	d) {\bf XMMU J021908.3-044731}, flat spectrum ($\Gamma=0.94_{-0.25}^{+0.23}$) object, still not identified.}  
              \label{eg_sp}%
\end{figure}

Fig.~\ref{nbins} shows the quality of the selected spectra as a
function of the 0.2-12 keV number of counts for the whole sample of 
sources (solid-line histogram) and for the objects identified as 
BLAGNs and NELGs (shaded histogram).  The vertical line
indicates the quality filter that we applied to select the spectra 
appropriate for the analysis.
In Fig.~\ref{eg_sp} we present, as an example, the X-ray spectra 
of some of the objects that we have analysed.
 
\section{Spectral Analysis}
\label{spectral_analysis}

The data were fitted with the XSPEC (version 11.2) software package. The
spectral analysis has been carried out fitting simultaneously MOS and
pn spectra, forcing the models to have the same parameters in both instruments,
including the normalisation (we do not expect significant calibration
mismatches between instruments in the energy interval considered and
for the quality level of our data).

We quantified the results of the spectral fits in terms of the
null hypothesis probability ${\rm P(\chi^2)}$ , i.e., the probability
that the observed spectrum is derived from the parent model under
scrutiny. We consider a model not good enough if ${\rm P(\chi^2)}$ is
less than 5\%.  

To test the significance of additional input parameters the F-test has
been applied with an adopted significance threshold of 95\%. As is
shown in Protassov et al. \cite{Protassov2002}, if the conditions to
use the F-test statistic are not satisfied, the false positive rate
might differ from the value expected for the selected confidence
level. One of the conditions that must be satisfied, is that the null
values of the additional parameters, should not be in the boundary of
the set of possible values of the parameter. We are using 
the F-test to calculate the significance of detection of 
absorption, and one of the possible values of the absorbing column is zero.
Hence, because one of the conditions for using the F-test is 
not satisfied, we do not know whether the fraction of spurious detections in 
our data differs from the 5\% expected. To calculate this number, we 
carried out simulations of unabsorbed spectra, and then we
calculated the number of cases where we detected absorption with
a confidence level above 95\%. We have detected absorption in $\sim$
2\% of the simulated spectra. As this fraction is not significantly
different from the expected value, we decided to be conservative, and
hence, we have continued with 5\% as the number of spurious detections,
or false positive rates.

\subsection{Single power law (model A)}
\label{modA}
We have started the analysis fitting a single power law to the background-subtracted spectra, 
hereafter model A.  This model has two free
parameters, the normalisation and the continuum slope
$\Gamma$. To account for the effect of the galactic neutral hydrogen
absorption along the line of sight, a fixed photoelectric absorption
component was included.  The Galaxy H column density values for each
field, ${\rm N_H^{Gal}}$, were extracted from the HI map of Dickey \&
Lockman (\cite{Lockman1990}).  For the selected acceptance level, 139
(12\%) objects had statistically unacceptable fits with this model,
the expected number for the selected confidence level being 57.  
Within the sub-samples of identified
sources, model A could not be accepted for 10 (7\%) BLAGNs and 9
(31\%) NELGs.

\subsection{Single power law with excess absorption (model B)}

Synthesis models of the XRB, based on AGN unification schemes
(Setti \& Woltjer \cite{Setti1989}, Madau, Ghisellini \& Fabian \cite{Madau1994}, 
Comastri et al. \cite{Comastri1995}, Pompilio, La Franca \& Matt \cite{Pompilio2000}, 
Gilli, Risaliti \& Salvati \cite{Gilli2001}) predict a large
population of highly absorbed AGNs. Hence we expect a significant fraction of
our objects to exhibit absorption in
excess of the Galactic value.  An indication of absorption 
is to measure a flat spectral slope with model A.
To study excess absorption, we fitted a model with two local
absorbers, one with the column density fixed at the Galactic value and
the other one as a free parameter ${\rm N_H^{obs}}$ (model B). Note
that this last term is essentially indistinguishable from an absorber
with column density $\sim {\rm N_H^{obs}\, (1+z)^{2.7}}$ 
(Barger et al \cite{Barger2002}; Longair \cite{Longair1992})
at the redshift of the source, $z$, since
absorption edges would not be detected in our highly grouped, moderate
signal-to-noise spectra.  This model has three free parameters, the
normalisation, the continuum spectral slope $\Gamma$, and the
equivalent absorbing column at $z=0$, ${\rm N_H^{obs}}$.  The top
panel in Fig.~\ref{h0_distr} compares the ${\rm P(\chi^2)}$
distributions obtained with model A with the results from model B for
the whole sample of sources. We see that the number of
unacceptable fits has been reduced by a factor of $\sim1.5$, implying
the presence of intrinsically absorbed objects.  

\subsection{Single power law with intrinsic absorption (model C)}

\begin{figure}
  \centering \hbox{
  \includegraphics[angle=90,width=8.5cm]{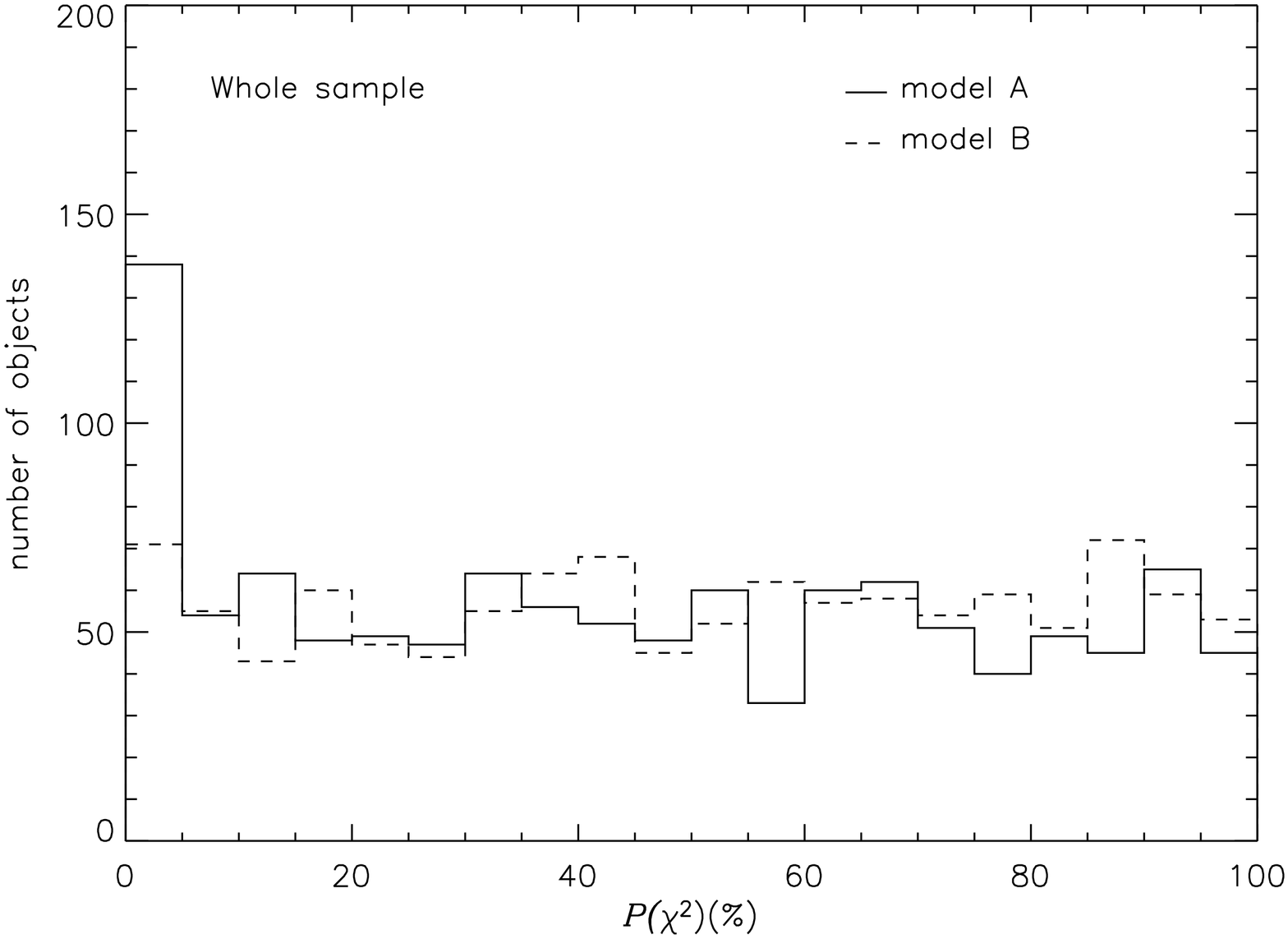}} \hbox{
  \includegraphics[angle=90,width=8.5cm]{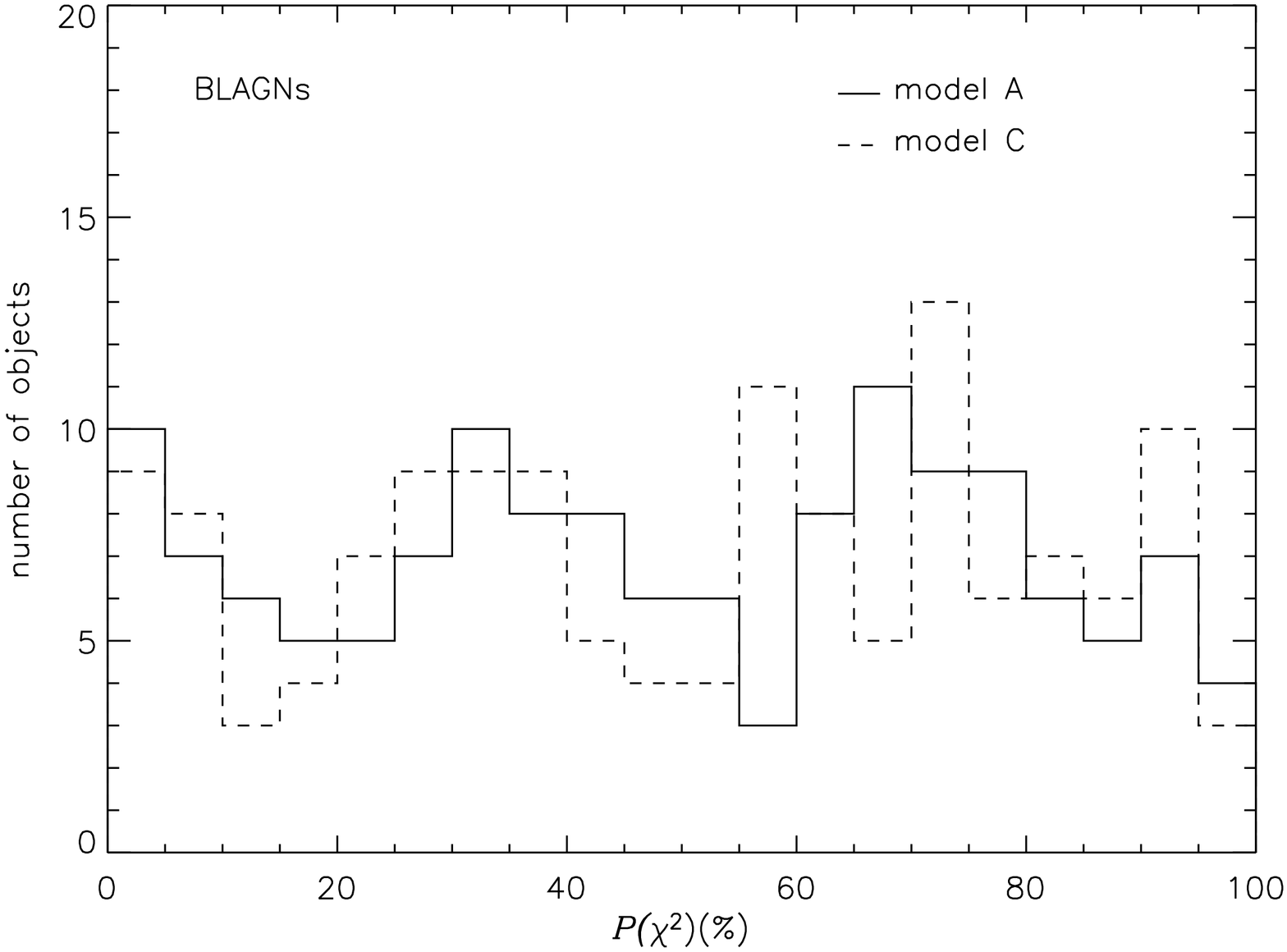}} \hbox{
  \includegraphics[angle=90,width=8.5cm]{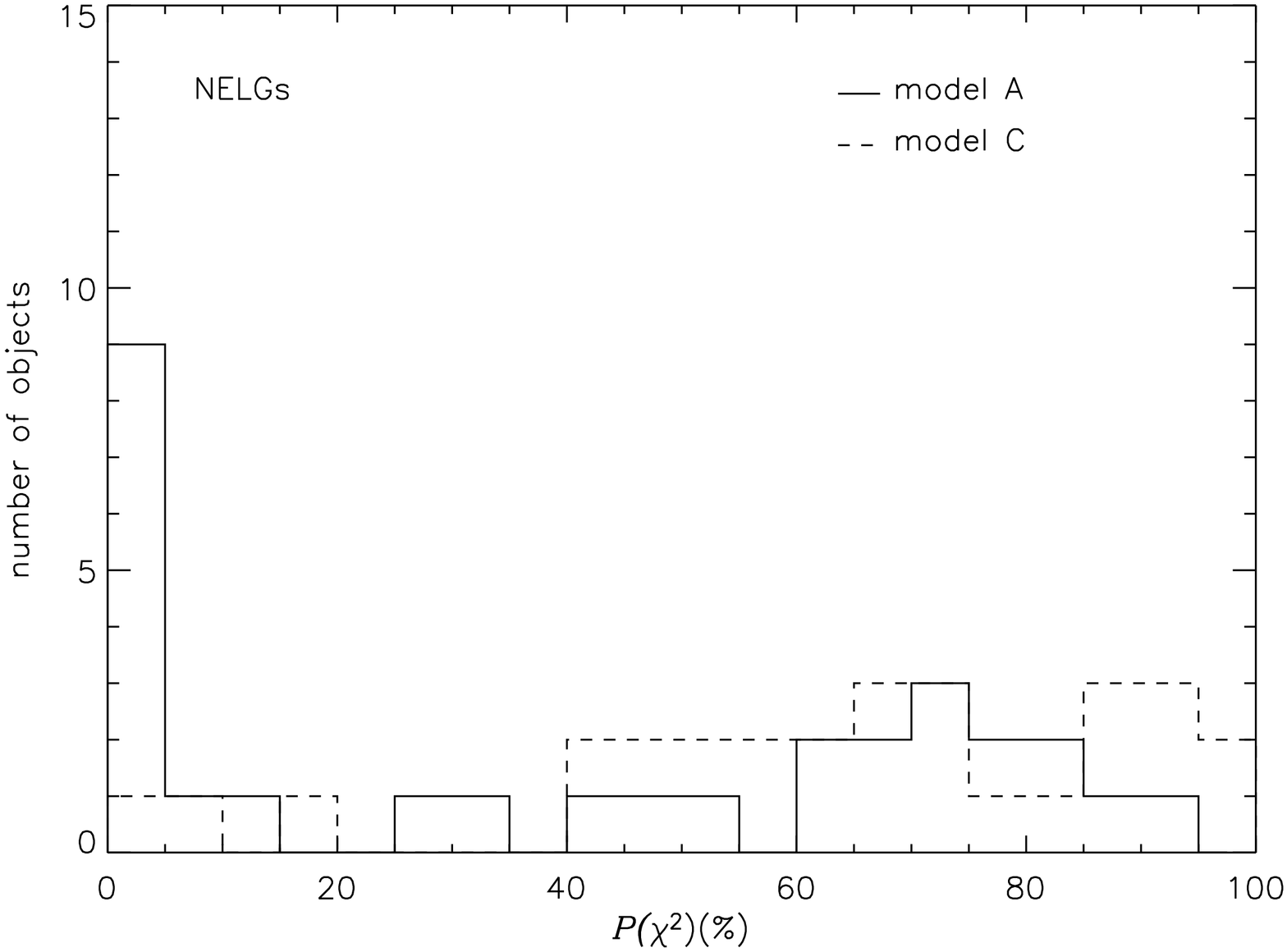}}
   \caption{Comparison of the null hypothesis probability
   distributions, $P(\chi^2)$, obtained from model A (simple power
   law, solid lines) and models B and C (observer frame absorbed power
   law and rest frame absorbed power law models, dashed lines). From
   top to bottom: whole sample of sources, BLAGNs and NELGs.}
   \label{h0_distr}%
\end{figure}

As an alternative, more physical parameterisation of the absorbing
column of the BLAGN, NELG and ALG,
we instead fitted photoelectric absorption at the redshift of the
source (model C).  The model has three free parameters, the
normalisation, the continuum spectral slope, $\Gamma$, and the
rest frame absorption, ${\rm N_H^{intr}}$.  Fig.~\ref{h0_distr}
compares the results of the fits obtained with model A and model C, for 
BLAGNs and NELGs. For a detailed description of the spectral properties of 
the 7 ALGs analysed, see Sec.~\ref{ALGs}.
In the case of the BLAGNs, we did not find a substantial improvement in
the overall quality of the fits, suggesting that the majority of the
BLAGNs do not require intrinsic absorption in excess of the Galactic
value. Therefore, for the statistically unacceptable fits, more
complicated models might be needed to reproduce the spectra of these
sources.  On the contrary, the observed significant improvement in the
quality of the NELG spectral fits obtained with model C, suggests that a
large fraction of NELGs are intrinsically absorbed.

\subsection{Absorbed power law and soft excess (model D)}

The soft excess has been shown to be a common feature of the X-ray spectra
of Seyfert 1 galaxies (Turner \& Pounds \cite{Turner1989}). The physical origin of
this emission is still unclear, since its determination depends on a
knowledge of the shape of the power law and the quantity of
absorption.  It has been interpreted as primary emission from the
accretion disc, gravitational energy released by viscosity in the disc, 
or as secondary radiation from the reprocessing of hard X-rays in the surface layers of the disc. 

We have searched for soft excess among the BLAGNs and NELGs.  We first
have fitted their spectra with a power law and a low energy black body 
component at the redshift of the source, absorbed by the
Galaxy. The soft excess emission, parameterised as a black body,
adds to the model two parameters, the black body 
temperature and normalisation.

We have compared the $\chi^2$ of the fit with the values obtained
with model A to search for the cases where the significance of a
$\chi^2$ improvement was above 95\%. From the 170 sources analysed
(141 BLAGNs and 29 NELGs) we found evidence of soft excess emission in 12
sources ($\sim$7\%), 10 BLAGNs and 2 NELGs.

As explained before, if soft excess is present in the
spectra of the objects, it must be modelled properly to prevent the 
continuum spectral slope $\Gamma$ and/or the intrinsic absorption 
${\rm N_H^{intr}}$ being
incorrectly determined.  Therefore we have analysed the X-ray spectra
of these 12 sources in more detail.  All the spectra have a large enough
number of bins, 35$<$bins$<$317, to use a model with five free
parameters: a power law and a low energy black body component, both
absorbed by the Galaxy (${\rm N_H^{Gal}}$; fixed) and by absorption (${\rm N_H^{intr}}$; free) at the redshift 
of the sources (hereafter, model D). Results from this specific exercise are
presented in Sect.~\ref{soft_excess}.

\section{Results of spectral fitting}
\label{sp_fitting}
\subsection{The continuum shape}
\label{continuum_shape}
Although we know that a large fraction of sources in our 
list are absorbed, we have started the study 
with the results that we obtained with model A 
to allow comparison with spectral analyses of 
data with low signal to noise. 
This model, as was explained before, does not take into account the effects of
absorption or soft excesses.

To calculate the objects average photon index, $\langle \Gamma \rangle$, we 
have used the standard formula for the weighted mean,
\[
\langle \Gamma\rangle=\sum\,P_i\times\,\Gamma_i
\]
where the weight, $P_i$, of each individual best fit value, $\Gamma_i$, is a function of the error 
obtained from the fit, $\sigma_i$, 
\[
P_i={1\,/\sigma_i^2 \over \sum\,( 1\,/ \sigma_i^2\,)}
\]
To calculate the uncertainty in $\langle \Gamma \rangle$ we have used the standard deviation 
(Bevington \cite{Bevington1992}) 
\[
\sigma^2={1 \over (\rm{N}-1)}\,\sum\,P_i\times( \Gamma_i-\langle \Gamma\rangle)^2
\]
that includes the measurement errors, $\sigma_i$, and   
the dispersion of each $\Gamma_i$ from the estimated value $\langle \Gamma \rangle$.

\begin{figure*}
    \hbox{
    \includegraphics[angle=90,width=8.5cm]{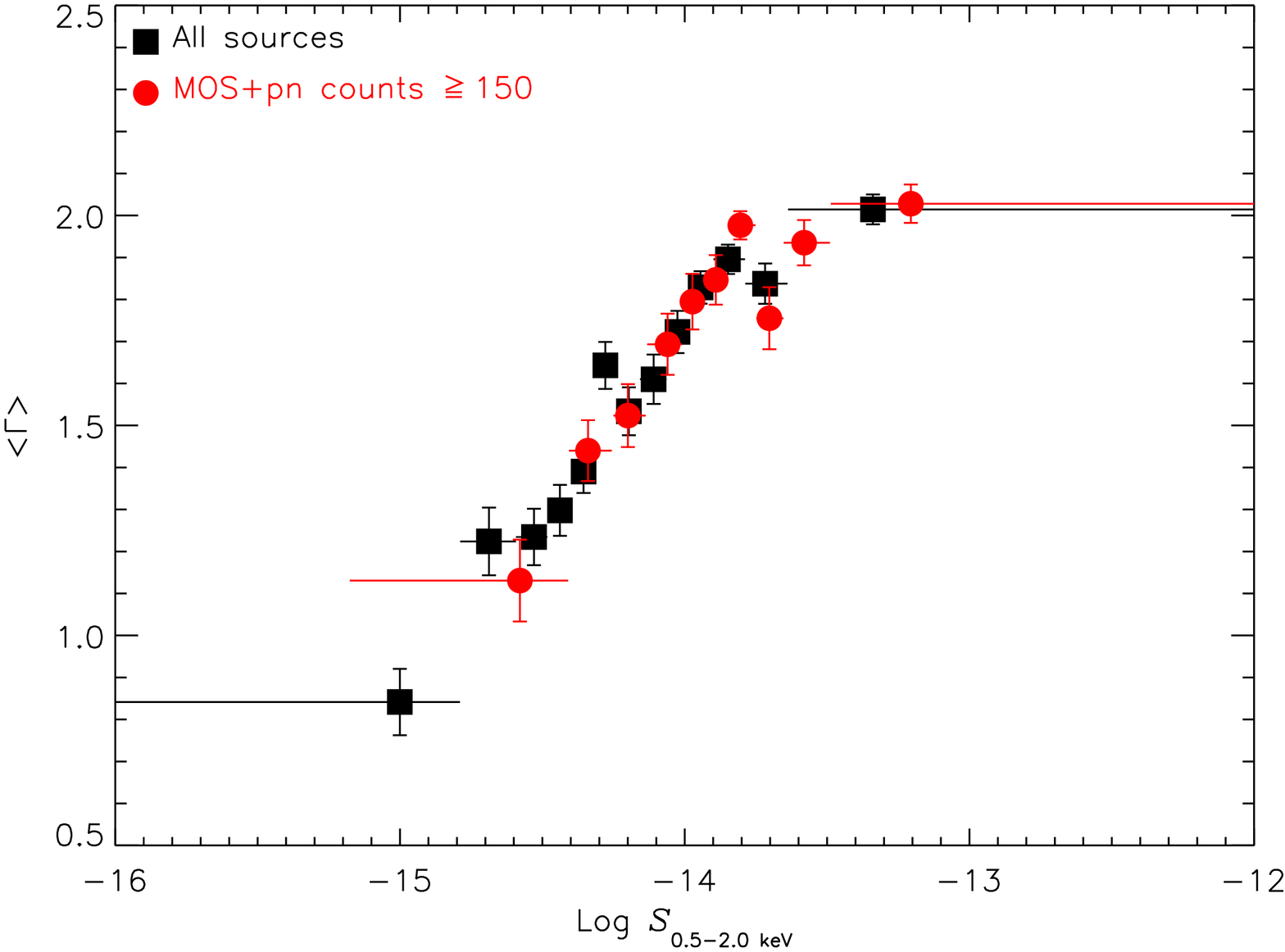}
    \includegraphics[angle=90,width=8.5cm]{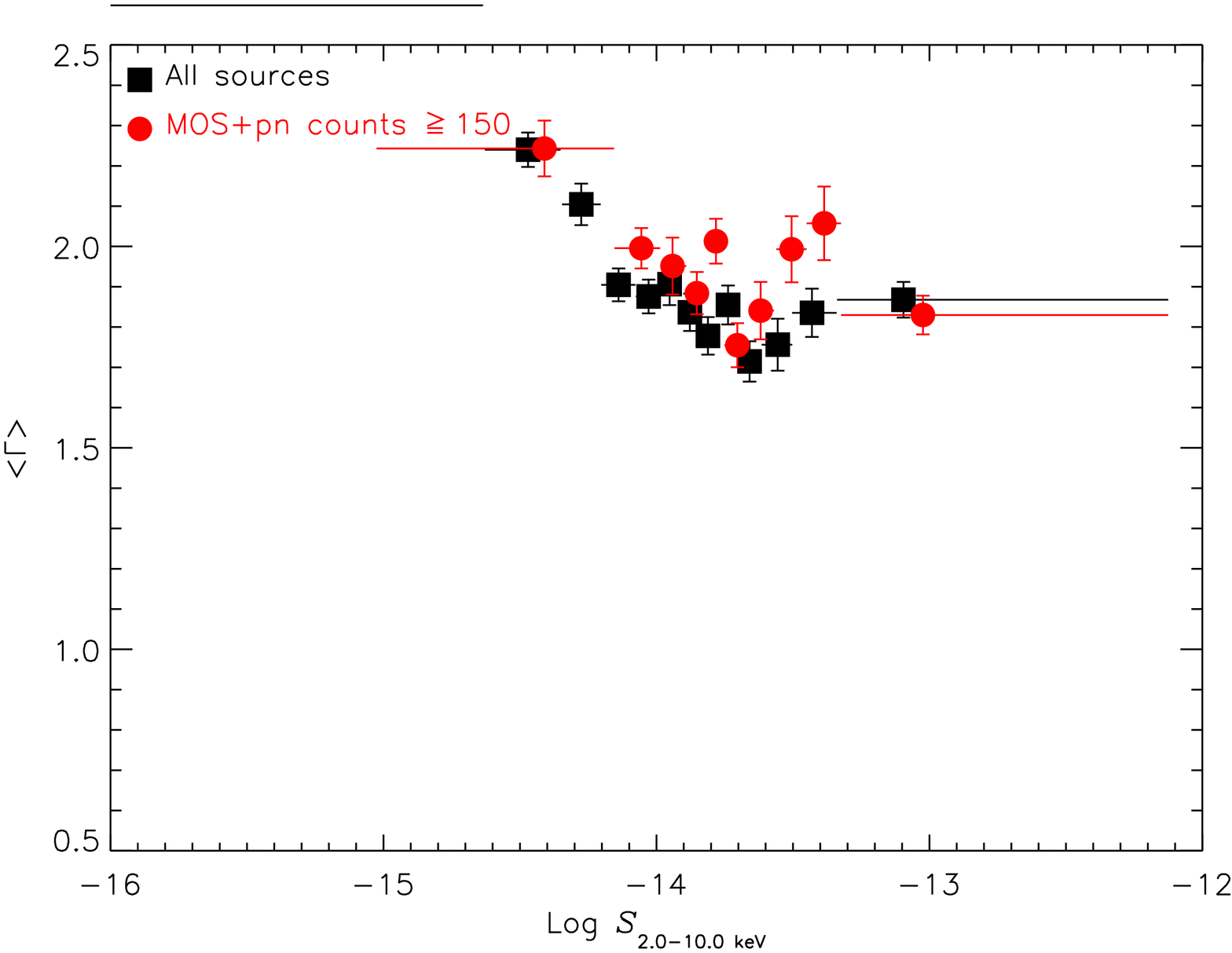}}
    \caption{Weighted photon index versus $S_{0.5-2\,{\rm keV}}$ and $S_{2-10\,{\rm keV}}$ 
      from model A (simple power law). The results 
      obtained for the objects with X-ray spectra with more than 150 counts
      (background subtracted) are included for comparison.
      Error bars correspond to 90\% confidence}
    \label{gamma_flux}
\end{figure*}

Following these definitions, we find our sources to have a continuum emission, from 0.2 to 12 keV,
with an average slope of $\langle\Gamma\rangle=1.86\pm0.02$ (the arithmetic mean being $1.70\pm0.02$).
To study the dependence of the objects average spectrum on the X-ray flux, we have calculated 
the values of $\langle \Gamma \rangle$ in bins of 
0.5-2 and 2-10 keV flux (each bin containing the same number of objects). 
Fig.~\ref{gamma_flux} shows the results that we have obtained.
We see that the average spectrum of the sources becomes harder with decreasing 0.5-2 keV flux, 
with $\langle\Gamma\rangle$ going from $2.01\pm0.03$ at $\sim4.4\times10^{-14}\,\mathrm{erg\,cm}^{-2}
\mathrm{s}^{-1}$ to $0.95\pm0.12$ at $\sim 10^{-15}\,\mathrm{erg\,cm}^{-2}\,\mathrm{s}^{-1}$. This trend 
\footnote{The same dependence, i.e., hardening of $\langle \Gamma \rangle$ at fainter 0.5-2 keV fluxes, 
is observed if $\langle \Gamma \rangle$ is calculated with the arithmetic mean, therefore the result is not an artifact of the weighted mean.} (harder
sources at fainter 0.5-2 keV fluxes) has been found in previous X-ray spectral analyses 
(e.g., Vikhlinin et al. \cite{Vikhlinin1995}, Mittaz et al. \cite{Mittaz1999}, Tozzi et al. \cite{Tozzi2001}, 
Mainieri et al. \cite{Mainieri2002}). Its origin 
is still not clear, although it could be explained as a rapid increase in the 
intrinsic absorption of the objects with decreasing flux, as it is predicted 
by some AGN synthesis models (see Gilli et al. \cite{Gilli2001})\\
It is not possible to see the same effect using the 2-10 keV fluxes, where $\Gamma$ seems to 
become softer at fainter fluxes (see Fig.~\ref{gamma_flux}). It is important to note that this 
softening of $\Gamma$ is not 
a real property of our objects. It comes from using different energy bands simultaneously 
in the source detection. As we go to 
fainter 2-10 keV fluxes, it is more difficult to detect sources with flat spectral 
slopes (hard objects) because their spectra peak at energies above $\sim$ 2 keV where we know the 
sensitivity of the X-ray detectors decreases rapidly.  Hence, the number of objects only detected in
 the soft band becomes more important at fainter 2-10 keV fluxes, specially 
below $\sim 2\times 10^{-14}\,{\rm erg\,cm^{-2}\,s^{-1}}$. Note that the last bin at the faintest fluxes (see Fig.~\ref{gamma_flux}) is located at a 
value of $\langle \Gamma \rangle$ of more than 2.5. The objects that contribute to this bin have spectra with very 
few counts and therefore the values of $\Gamma$ are poorly constrained. 

We have studied in more detail whether the observed hardening 
of $\Gamma$ with decreasing soft flux could be affected by any bias
introduced during our analysis. 
We first have checked that the same result is obtained when only sources with more than 
150 counts (circles in Fig.~\ref{gamma_flux}) are used. Therefore, the sources with 
poor spectral quality are not introducing any bias in our result. 

\begin{figure}
    \hbox{
    \includegraphics[angle=90,width=8.5cm]{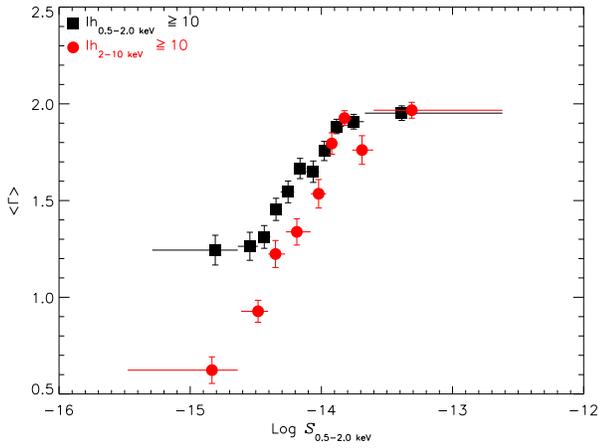}}
    \caption{Weighted photon index versus $S_{0.5-2\,{\rm keV}}$ from model A (simple power law)
	 for sources with detection likelihood $\ge$10 in the 0.5-2 keV band (squares) and 
	 for sources with detection likelihood $\ge$10 in the 2-10 keV band (circles).
	Error bars correspond to 90\% confidence}
    \label{gamma_flux_soft}
\end{figure}

We expect the majority of the observations that we 
have used, not to be deep enough as to detect sources with a 
 0.5-2 keV flux of ${\rm \sim 10^{-15}\,erg\,cm^{-2}\,s^{-1}}$ in the soft band. 
Hence, it could be possible that our $\langle \Gamma \rangle$ becomes harder with decreasing 
flux partly due to the contribution to each bin from bright hard sources 
not detected in the soft band.
This can be checked by repeating the plot of Fig.~\ref{gamma_flux} with 
those sources selected in the soft or hard bands (detection likelihoods above 10 on each band). 
The results, plotted in Fig.~\ref{gamma_flux_soft}, show a very similar
dependence of $\Gamma$ with flux. Only the 
$\langle \Gamma \rangle$ of the bin at the faintest soft fluxes in Fig.~\ref{gamma_flux}, 
could be affected or even dominated by the contribution of bright hard sources not detected in the soft band. 
It is interesting to note that the hardening of $\Gamma$ is more important for objects detected in the 2-10 keV 
band (see Fig.~\ref{gamma_flux_soft}). If absorption is producing the hardening of $\Gamma$, this result suggests that we are seeing the same 
population of objects in the soft and hard bands, the only difference being that the objects detected in the 2-10 keV band are more absorbed on average.

Finally we have used the source detection efficiency function, W($\Gamma,{\it S}$) (see Appendix ~\ref{apendix_A}), to correct our results for all the biases of our study. The 
de-biased average values of $\Gamma$ at each bin of flux 
were obtained weighting each individual value, $\Gamma_i$ with the function:
\[
P_i={1\,/(\sigma_i^2\times W(\Gamma_i,{\it S_i})) \over \sum\, 1\,/ (\sigma_i^2\times W(\Gamma_i,{\it S_i}))}
\]
where the sum is calculated with all the sources falling in the specified bin in the $\Gamma$, {\it S} plane (for a detailed 
explanation see Appendix ~\ref{apendix_A}).

\begin{figure*}
    \hbox{
    \includegraphics[angle=90,width=8.5cm]{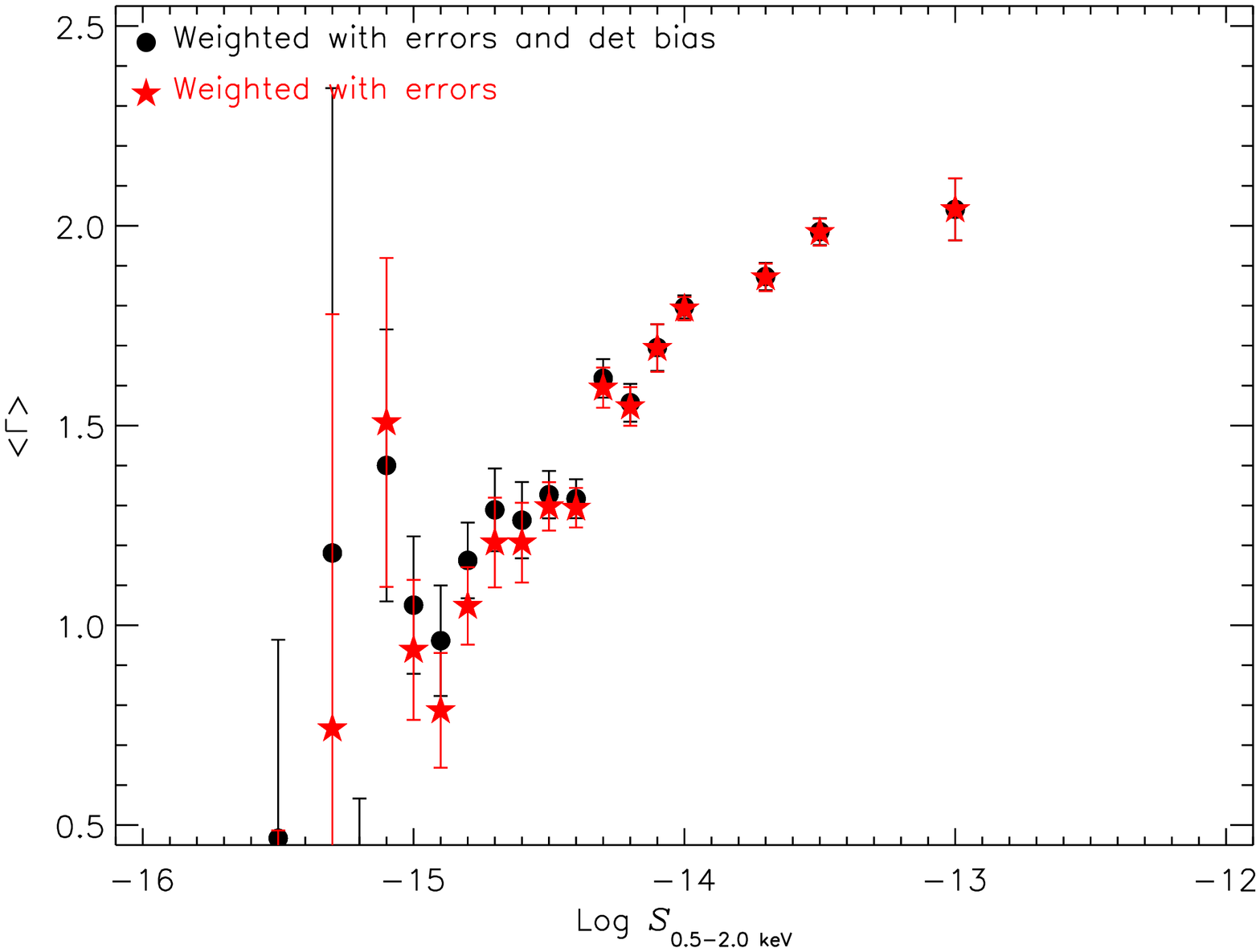}
    \includegraphics[angle=90,width=8.5cm]{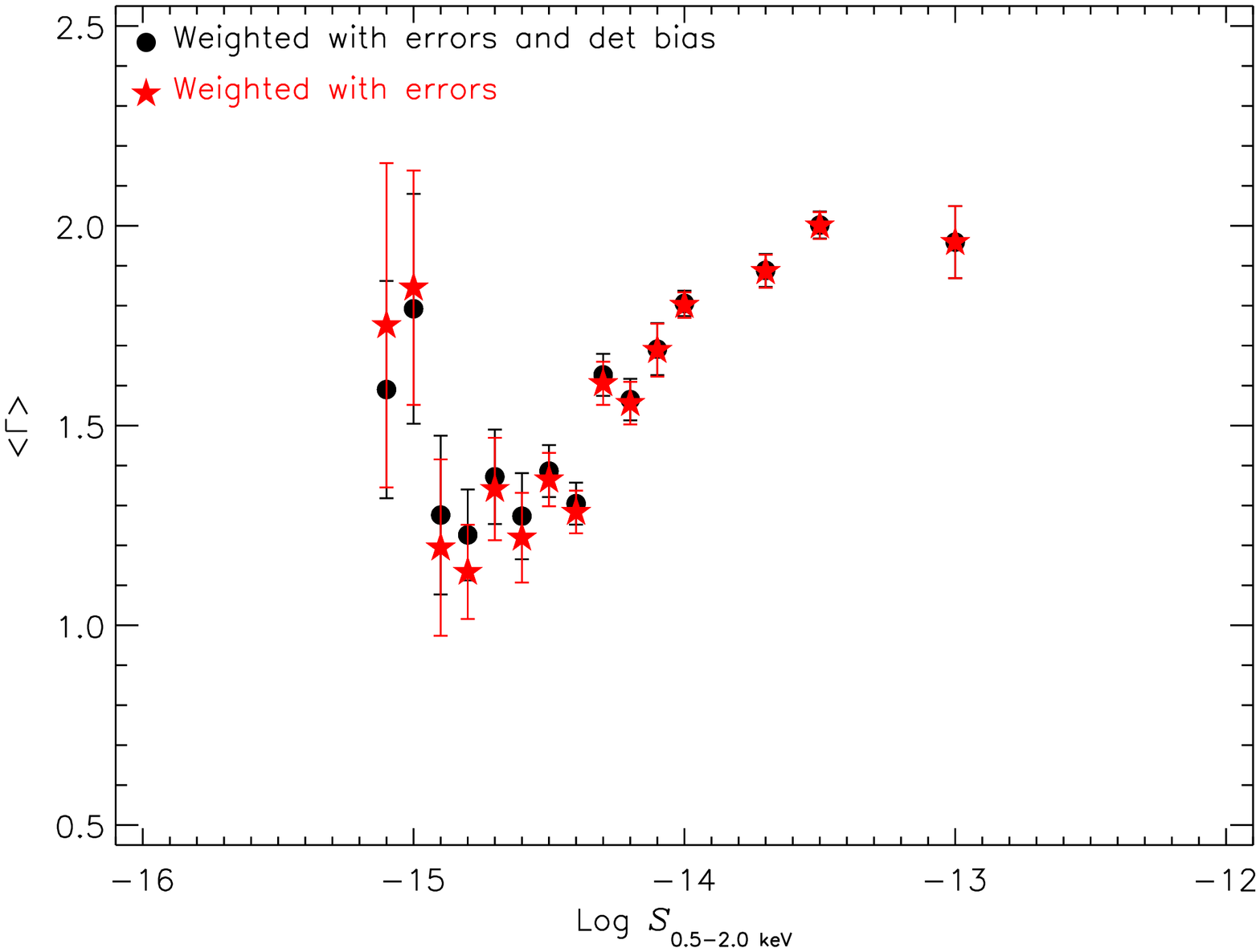}}
    \caption{Comparison of the weighted photon index versus $S_{0.5-2\,{\rm keV}}$ from 
      model A (simple power law) with the values obtained after correcting for the source 
      list biases. Left for the whole sample of objects and right for the sub-sample 
	of sources with detection likelihood $\ge$10 in the 0.5-2 keV band.
	Error bars correspond to 90\% confidence}
    \label{gamma_flux_debiased}
\end{figure*}

The results are shown in Fig.~\ref{gamma_flux_debiased}, the left plot for the whole sample of 
sources, and the right plot for the sources detected in the soft band. For comparison we have 
calculated the $\langle \Gamma \rangle$ values weighted with the fitting errors in the same flux bins 
(stars in the plots). We still see the hardening of $\langle \Gamma \rangle$ with the 0.5-2 keV 
flux, so this dependence 
is indeed a real property of the population of the sources analysed.

\subsection{Excess Absorption}
\label{intrinsic_absorption}

The fit to the X-ray spectra with a single power law (model A) yields 
a large number of sources with statistically
unacceptable fits. Models B and C reduce significantly the number of
unacceptable fits as it is shown in Fig.~\ref{h0_distr}, suggesting that
a substantial fraction of our sources could be X-ray absorbed, in
agreement with the predictions of the standard synthesis models of the
XRB (Comastri et al. \cite{Comastri1995}, Gilli et al
\cite{Gilli2001}). 

To search for absorbed objects we used the
F-test, measuring the improvement in the $\chi^2$ of the fits after
adding absorption to the model. A 95\% confidence cutoff is adopted
for absorption to be detected. In Table~\ref{tab2}, we show the total number 
of objects of different types, the number
of absorbed objects of that type, and the fraction of absorbed objects of each
type with its uncertainty and significance.
  
For a sample of $N$ objects, of which $n$ are absorbed, the Bayesian
posterior probability of the fraction of absorbed objects is $P(f)\propto
(f+f_{spu})^n(1-f-f_{spu})^{N-n}$ (from the binomial distribution), where
$f_{spu}$ is the fraction of spurious detections (for our sources 5\%).
We used the mode of {\it P(f)} to estimate the fraction (corrected from spurious detections) 
of absorbed objects within each class of sources. The errors in the fractions were calculated 
integrating {\it P(f)} from the mode value in the two directions until we obtained 
45\% of the probability (90\% errors). 
For every ``class'' of sources, we measured the confidence of detection of absorption 
integrating the {\it P(f)} distributions from {\it f}=1 down to the value where the integral 
is equal to 0.9973 (or 3$\sigma$). These values, listed in the last column of 
Table~\ref{tab2}, indicate the fraction of absorbed objects that we have detected within each 
class of sources with a significance of more than 3$\sigma$. For example, we know with a confidence 
of more than 3$\sigma$ that at least $\sim$ 1\% of BLAGNs and $\sim$ 17\% of NELGs are absorbed. 

It appears that the fraction of absorbed objects varies substantially
among different classes of object, with the NELGs having the highest
fraction. We have used the method described in Stevens et al. 
\cite{Stevens2005} to study whether our samples of BLAGNs and NELGs have 
different fractions of absorbed objects. Assuming that BLAGNs and NELGs 
come from the same parent population, the fraction of absorbed objects is 
\[
P(f)\propto(f+f_{spu})^{(n+m)}(1-f-f_{spu})^{(N+M-n-m)}
\]
where {\it M} and {\it N} are the number of sources in the two samples and {\it m} and {\it n} 
the number of detections on each sample.

The probability of detection of {\it i} objects in the first sample and {\it j} in the second one is 
given by

\[
P(i,j;N,M)={M\choose j}{N\choose i}\,\int_0^1\,df\,P(f)\,(f+f_{spu})^{(i+j)}
\]
\[
\,\,\,\,\,\,\,\,\,\,\,\,\,\,\,\,\,\,\,\,\,\,\,\,\,\,\,\,\,\,\,\,\,\,\,\,\times\,\,\,(1-f-f_{spu})^{(N+M-i-j)}
\]
The fractions of absorbed objects in BLAGNs and NELGs will be different if 
the probability of detecting $\le m$ in the first sample and $\ge n$ in the second 
one is low, i.e. if $P(\ge n, \le m;N,M)=\sum_{j=0}^{M}\sum_{i=n}^{N}P(i,j;N,M)$ is low.
We find that $P(m\le16,n\ge13;141,29)=0.000013$. Therefore we see that the fraction of absorbed 
objects in BLAGNs and NELGs is different with a significance of more than 3$\sigma$. 
We cannot distinguish the fraction of absorbed objects among NELGs and ALGs 
(the probability of them being different is not significant). 

\begin{figure}
    \resizebox{\hsize}{!}{
    \includegraphics[angle=90,width=\textwidth]{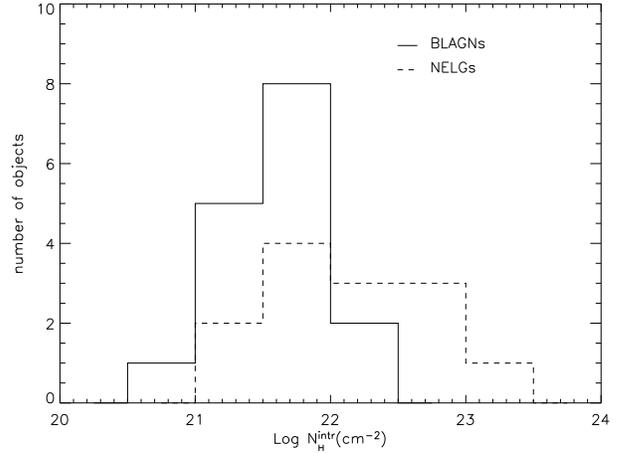}}
    \caption{Distributions of intrinsic absorption for BLAGNs (solid
    histogram) and NELGs (dashed histogram).
    } \label{nh_in}
\end{figure}

Fig.~\ref{nh_in} shows the intrinsic ${\rm N_H^{intr}}$ distributions
measured for the absorbed BLAGNs and NELGs. BLAGNs show column
densities $4.5\times10^{20}<{\rm N_H^{intr}}<2.3\times10^{22}\,{\rm
cm}^{-2}$ while NELGs seem to cover a much wider range of
absorption, with $1.1\times10^{21}<{\rm
N_H^{intr}}<1.7\times10^{23}\,\mathrm{cm}^{-2}$.  However, the small
number of objects does not allow us to find a significantly different
(in terms of the Kolmogorov-Smirnov test) distribution of absorbing
columns between BLAGNs and NELGs.

\begin{figure}
  \hbox{
  \includegraphics[angle=90,width=8.5cm]{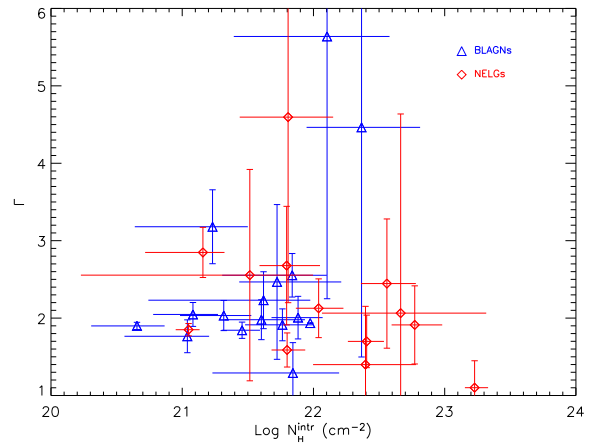}}
  \caption{Photon index versus intrinsic absorption for
  absorbed (F-test significance $\ge95\%$) BLAGNs (triangles) and NELGs (diamonds).
	}
\label{gamma_nh}
\end{figure}

To be confident that the detected intrinsic absorptions among 
identified sources were not affected by the quality of their spectra,
we have plotted $\Gamma$ versus ${\rm N_H^{intr}}$ to search for 
any correlation between the two spectral
parameters. Such a correlation would be expected if the detections of
absorption were spurious and linked to very large values of
$\Gamma$. The results are shown in Fig.~\ref{gamma_nh}. 
The plot shows no apparent correlation between 
$\Gamma$ and ${\rm N_H^{intr}}$.

A Kolmogorov-Smirnov comparison between the redshift
distributions of unabsorbed and absorbed BLAGNs does not indicate a
significant difference between the two. The same result is obtained when comparing the 2-10 keV 
luminosity distributions of absorbed and unabsorbed BLAGNs and NELGs 
(we use the 2-10 keV energy band because it is less affected by absorption). 
We conclude that X-ray absorption does not occur preferentially at any 
particular redshift (in BLAGNs) or X-ray luminosity.

Finally, we have tested whether the lack of optical broad emission lines
in unabsorbed NELGs can be explained in terms of AGN/host galaxy contrast 
(Lumsden et al. \cite{Lumsden2001}, Moran et al. \cite{Moran2002}, 
Page et al. \cite{Page2003}, Severgnini et al. \cite{Severgnini2003}) . To this 
end, we have compared the 2-10 keV luminosity distributions of two sub-samples of 
unabsorbed BLAGNs and NELGs detected with redshifts ranging from 0.17 to 0.7, where 
the redshift distributions of BLAGNs and NELGs overlap (see Fig.~\ref{zdist}).
The Kolmogorov-Smirnov significance of the two distributions being different 
is $\sim$ 99\%. Furthermore, the average luminosities of the selected BLAGNs and NELGs 
were found to be 43.7$\pm$0.16 and 43.2$\pm$0.11 (in log units).
Therefore, we have seen that our unabsorbed NELGs are on average significantly less 
luminous than unabsorbed BLAGNs at similar redshifts, and hence, our data are consistent with the explanation of 
non detection of broad optical lines in unabsorbed NELGs as an AGN/ host-galaxy contrast effect.

\begin{table}
\caption{Fractions of F-test absorbed objects and significances.} 
\begin{tabular}{l c c c c c c}
\hline
 Type & ${\rm N_{tot}}$ & ${\rm N_{abs}}$ & ${\it f_-^+}^{\mathrm{a}}$ & ${\rm F^{\mathrm{b}}}$\\
\hline
	  {\rm All} & 1137 & 245 & $0.17\,_{-0.02}^{+0.02}$ & 0.1335 \\
	  {\rm BLAGNs} & 141 & 16 & $0.06\,_{-0.04}^{+0.05}$ & 0.0089\\
	  {\rm NELGs} & 29 & 13 & $0.40\,_{-0.14}^{+0.15}$ & 0.1740\\
	  {\rm ALG} & 7 & 3 & $0.40\,_{-0.20}^{+0.20}$  & 0.0400\\  
\hline
\end{tabular}

\begin{list}{}{}
\item[$^{\mathrm{a}}$] The f values are the mode of the function {\it P(f)} 
  (see Sec.~\ref{intrinsic_absorption} for details). The errors are 90$\%$
\item[$^{\mathrm{b}}$] F is where {\it P}({\it f} $\ge$ F)=0.9973 (3$\sigma$)
\end{list}
\label{tab2}
\end{table}

\subsection{Soft excess}
\label{soft_excess}
We have previously shown that $\sim$7\% of the sources 
identified as BLAGNs or NELGs exhibit significant soft excess 
emission. To measure the broad
band spectral parameters of these sources, and in particular the
underlying power law index $\Gamma$ we have used model D, as described
in detail in Sec.~\ref{spectral_analysis}.

   \begin{table*}
   \begin{minipage}{100mm} \caption[]{Spectral parameters of sources
   with a soft excess component detected} \label{tab4} $$
          \begin{array}{c c c c c c c c c c c c} \hline
          \noalign{\smallskip}
            \mathrm{Name\,^{{\rm a}}} & \mathrm{Class} & \mathrm{redshift} &
            \Gamma_-^+ & \mathrm{N_{H}^{intr}}\,^{\mathrm{b}} & \mathrm{kT_-^+}
            & \mathrm{{\it S}_{0.5-2}} & \mathrm{L_{0.5-2}} & \mathrm{model
            \,C}\,^{\mathrm{c}} & \mathrm{model\,D}\,^{\mathrm{d}}\\ & &
            & & (10^{20}\,\mathrm{cm}^{-2}) & \mathrm{(keV)} & (\%) &
            (\%) & \\ \noalign{\smallskip} \hline \hline
            \noalign{\smallskip} \mathrm{XMMU\,J001831.7+162924} &
            \mathrm{BLAGN} & 0.551 &
            2.35_{-0.18}^{+0.17}&4.15_{-2.37}^{+11.25} &
            0.186_{-0.126}^{+0.259}&15.95 & 23.39 &\mathrm{Not\,\,\,abs.} &
            \mathrm{abs.}\\ \mathrm{XMMU\,J021808.2-045844} &
            \mathrm{BLAGN} & 0.712 &2.08_{-0.05}^{+0.04}&<10^{20} &
            0.142_{-0.111}^{+0.164}&6.55 & 19.49 & \mathrm{Not\,\,\,abs.} &
            \mathrm{Not\,\,\,abs.} \\ \mathrm{XMMU\,J021830.5-045622} &
            \mathrm{BLAGN} & 1.430 &2.28_{-0.18}^{+0.10}&<10^{20}&
            0.129_{-0.082}^{+0.223}&1.89 & 30.08 & \mathrm{Not\,\,\,abs.} &
            \mathrm{Not\,\,\,abs.}\\ \mathrm{XMMU\,J061728.6+710600} &
            \mathrm{BLAGN} & 0.219 & 1.68_{-0.37}^{+0.56}&<10^{20} &
            0.214_{-0.111}^{+0.253}&39.54 & 44.57 & \mathrm{Not\,\,\,abs.} &
            \mathrm{Not\,\,\,abs.} \\ \mathrm{XMMU\,J061730.9+705955} &
            \mathrm{NELG} & 0.200 &2.02_{-1.00}^{+1.13}&<10^{20} &
            0.207_{-0.142}^{+0.272}&44.14 & 49.31 & \mathrm{abs.} &
            \mathrm{Not\,\,\,abs.}\\ \mathrm{XMMU\,J074311.9+742935} &
            \mathrm{BLAGN} & 0.332 &2.05_{-0.05}^{+0.03} &<10^{20} &
            0.119_{-0.106}^{+0.130}&12.22 & 22.12 & \mathrm{Not\,\,\,abs.} &
            \mathrm{Not\,\,\,abs.} \\ \mathrm{XMMU\,J12226.1+752616} &
            \mathrm{NELG} & 0.238 &1.82_{-0.10}^{+0.14}
            &11.17_{-2.30}^{+2.43} & 0.146_{-0.112}^{+0.208}&12.30 & 22.72
            &\mathrm{abs.} & \mathrm{abs.} \\
            \mathrm{XMMU\,J13316.2+241326} & \mathrm{BLAGN} & 0.174&
            1.83_{-0.26}^{+0.45} &<10^{20} & 0.085_{-0.064}^{+0.116}&14.71
            & 25.20 & \mathrm{Not\,\,\,abs.} & \mathrm{Not\,\,\,abs.}\\
            \mathrm{XMMU\,J212904.5-154448} & \mathrm{BLAGN} & 0.431
            &1.29_{-0.36}^{+0.27}&14.21_{-5.81}^{+32.59} &
            0.205_{-0.142}^{+0.251}&54.55 & 72.74 & \mathrm{Not\,\,\,abs.} &
            \mathrm{abs.} \\ \mathrm{XMMU\,J221515.1-173224} &
            \mathrm{BLAGN} & 1.165 & 2.06_{-0.35}^{+0.37}&<10^{20} &
            0.212_{-0.192}^{+0.233}&24.75 & 49.69 & \mathrm{Not\,\,\,abs.} &
            \mathrm{Not\,\,\,abs.} \\ \mathrm{XMMU\,J221523.6-174321} &
            \mathrm{BLAGN} & 0.956 &2.15_{-0.23}^{+0.17} &<10^{20} &
            0.112_{-0.021}^{+0.234}&1.86 & 19.16 & \mathrm{Not\,\,\,abs.} &
            \mathrm{Not\,\,\,abs.} \\ \mathrm{XMMU\,J222823.6-051306} &
            \mathrm{BLAGN} & 0.758
            &1.52_{-0.15}^{+0.14}&15.61_{-7.20}^{+12.39} &
            0.262_{-0.171}^{+0.379}&35.78 & 54.55 & \mathrm{Not\,\,\,abs.} &
            \mathrm{abs.} \\ \noalign{\smallskip} \hline \end{array}
            $$
\begin{list}{}{}
\item[$^{\mathrm{a}}$] {\it XMM-Newton} source name 
\item[$^{\mathrm{b}}$] Intrinsic absorption measured with model D
\item[$^{\mathrm{c}}$] Detection of intrinsic absorption with model C
(F-test comparison of the $\chi^2$ from model A and model C)
\item[$^{\mathrm{d}}$] Detection of intrinsic absorption with model D (F-test
comparison of the $\chi^2$ from model A and model D)
\end{list}
\end{minipage}
   \end{table*}

The results of the fits are listed in Table~\ref{tab4}, where for each source
we have included its IAU-style {\it XMM-Newton} source name, the
optical classification, redshift, and the spectral parameters measured
with model D. We have also included the soft excess contributions to
the flux and luminosity in the soft band and the F-test results
obtained from the comparison of $\chi^2$ from model A to $\chi^2$ 
from model C and model D. We now discuss some particular
sources in detail:

{\bf XMMU J001831.7+162924, XMMU J222823.6-051306, XMMU
J212904.5-154448}: we obtained an improvement in the spectral fits
with model C, but the F-test significance was in all cases below 
95\%, therefore the sources were classified as
unabsorbed.  A proper modelling of the soft excess emission with model
D allowed us to find significant absorption above the Galactic
value in all the sources.

{\bf XMMU J221515.1-173224}: we obtained an unusually steep spectral
slope with model A, $\Gamma=2.67_{-0.06}^{+0.05}$. The quality of the 
fit did not improve significantly with model C. With model D we obtained a
significant improvement in the fit ($>99.99\%$). This model did not
require absorption in excess of the Galactic value . The measured
temperature of the soft excess emission was kT=$0.05_{-0.01}^{+0.03}$,
below the typical values found for the rest of the sources. The
fitted continuum was still significantly steep, with
$\Gamma=2.60_{-0.11}^{+0.04}$.We tried to find a more typical value of
$\Gamma$ fitting the continuum slope at high energies ($>2$ keV) with
model A. We obtained a value of $\Gamma$ significantly lower,
$2.06_{-0.35}^{+0.37}$ and in agreement with the average spectral
slopes observed in unabsorbed AGN. With $\Gamma$ fixed to the above
value the fit did not improve with model C, but with model D we still
obtained an F-test significance of $>99.99\%$. The model did not require
absorption in excess of the Galactic value, therefore we believe the
improvement is due entirely to the soft excess component. We adopted
this model as the best fit model for the source although the $\chi^2$
of the fit was slightly worse that the value obtained with $\Gamma$
fully variable.

{\bf XMMU J061730.9+705955}: we found a significant column density
with model C, hence the source was classified as absorbed. However we
obtained a very steep spectral slope,
$\Gamma=2.87_{-0.31}^{+0.35}$. With model D we obtained a better fit
of the data and more typical values for the spectral parameters,
although model D did not require intrinsic absorption. We have
chosen model D in preference to model C for this source.
 
It is interesting to note that for some of the sources that we have 
analysed, we
found very flat spectral slopes with model A, but they all required
significant quantities of soft excess emission.  This is the case for
{\bf XMMU J12226.1+752616} and {\bf XMMU J222823.6-051306}, with measured model A
spectral slopes of $\Gamma=1.52_{-0.04}^{+0.04}$ and
$\Gamma=1.78_{-0.11}^{+0.11}$ respectively.

\subsection{Results from best fit model}
\label{best_fit_model}

\begin{figure}
  \centering \hbox{
  \includegraphics[angle=90,width=8.5cm]{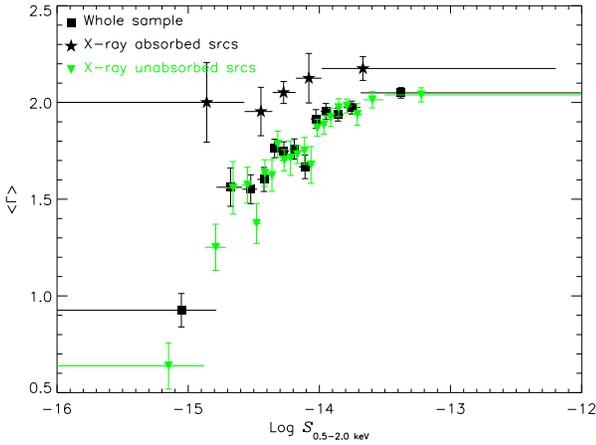}}
  \caption{$\langle\Gamma\rangle$ versus $S_{0.5-2\,{\rm keV}}$
  obtained from the best fit model of each object. Squares denote
  the whole sample; stars absorbed sources (F-test$\ge95.0\%$) 
  and triangles unabsorbed sources (F-test$<95.0\%$).}  
\label{gamma_flux_best_fit}
\end{figure}

\begin{figure}
  \centering \hbox{
  \includegraphics[angle=90,width=8.5cm]{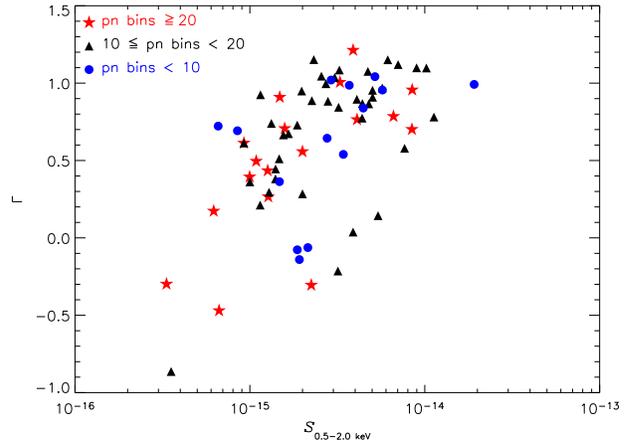}} \caption{
  $\Gamma$ versus $S_{0.5-2\,{\rm keV}}$ for objects with flat
  spectrum ($\Gamma$ positive error bar $<$1.5).
  We have grouped the objects in terms of the number of bins in their pn spectrum.}
  \label{gamma_bins}
\end{figure}

To test whether the spectral hardening observed using model A 
is due to an increase in intrinsic absorption at the faintest fluxes, we
compared $\Gamma$ (from the best fit model in terms of the F-test results) 
to the soft X-ray flux for each source. The best fit
model can be model A or model B for the unidentified sources, and
model A or model C for the identified sources (or model D for 
BLAGNs and NELGs when soft excess emission was detected). Using the best
fit model for each object we obtained a value of $\langle\Gamma\rangle=1.96\pm0.01$ 
(the unweighted value was $2.08\pm0.04$). The dependence of $\langle \Gamma \rangle$
with the soft flux is plotted in
Fig.~\ref{gamma_flux_best_fit}, where we see that:
\begin{enumerate}
\item The average continuum of absorbed sources is significantly softer than the 
continuum of unabsorbed sources, especially at the faintest fluxes,
with $\langle\Gamma\rangle=2.15\pm0.04$ (the unweighted value being $3.12\pm0.12$\footnote{
The unweighted mean is very high and different from the value that we 
obtained for the weighted mean, because there are a number of absorbed objects 
for which $\Gamma>$ 2.5 but with large errors bars.})
  for the absorbed sources and
$\langle\Gamma\rangle=1.94\pm0.01$ (the unweighted value being $1.79\pm0.02$)
 for the unabsorbed sources.
 \item There is no hardening of the average spectral slope for 
the absorbed sources. The hardening of $\Gamma$ seen in these objects  
when fitted with the SPL model is likely to be an effect of absorption only.
\item For the sources where absorption was not detected in terms of the F-test, we 
still observe a significant hardening of the average spectral
slope with the X-ray flux.
\end{enumerate}

Since the sources where absorption is detected do not
exhibit any significant variation of the underlying spectral index
with flux, does this hold true for the remaining sources? 
We first wanted to check whether any effect
could lead to an apparent flattening of the spectral shape towards
fainter fluxes when absorption is included in the fitting model.  
To this goal, we have generated a number of fake
spectra with an underlying power law $\Gamma=2$ at various fluxes
spanning those of our sample, and different amounts of absorption at
$z=0$. 

We have then first fitted these spectra with models A and B, and then 
we have compared the $\chi^2$ of the fits with the F-test. We found a 
range of low values of ${\rm N_H^{obs}}$ where absorption was not 
significantly detected when model A was replaced by
model B (F-test $<$ 95\%).  In these cases, the best fit value of $\Gamma$ (
i.e., the value obtained with model A) was indeed lower
than the input one, and it can reach values in the range
$\Gamma\sim$1-2.

A second result of this exercise is, if we hold the 
absorption in the simulated spectra but vary the 
flux, at fainter fluxes the typical values of $\Gamma$ fitted are
 marginally harder than at higher fluxes (and then fitted values of ${\rm N_H}$ tend to 
be lower than the input ones). We believe this effect to be due to a
degradation of the spectral resolution of the X-ray spectrum, produced
by the necessary grouping in very wide bands to achieve the desired
statistics (10 counts/bin).  This leads, in particular, to a smoothing
of the model count rate function (which is the effective area times
the input spectrum, convolved with the redistribution matrix) in the
whole spectral range, but in particular in the vicinity of its peak at
soft energies where most of the counts are detected.  This produces
an undesired slight hardening of the spectrum.

However, this effect is unlikely to produce the hardening of the
unabsorbed sources that we see.  The reason is that in this case we
should see the lowest quality spectra (i.e., those with the smaller
number of bins) being more prone to this effect, but this is not the case. 
Fig.~\ref{gamma_bins} shows the sources that we detect in
the real sample with flat spectra (best fit 
$\Gamma$ error bar upper limit below 1.5), 
grouped in terms of the
number of bins in their EPIC-pn spectra. We clearly see that the
hardening is not correlated with the quality of the spectra, and
therefore the above effect does not dominate. 

We conclude that the hardening towards fainter fluxes is real. This still 
leaves two options for its origin: intrinsic flattening
of the power law or undetected low level absorption. Although the
current data do not allow us to distinguish between them, there is
nothing inconsistent with the simplest hypothesis which is that these
sources are absorbed at a level which is undetectable in the current
data.This is also supported by the results of Fig.~\ref{gamma_flux} (right plot)
where we do not see hardening of $\Gamma$ with the 2-10 keV 
flux down to $\sim 2\times 10^{-14}{\rm erg\,cm^{-2}\,s^{-1}}$. If absorption is producing the 
hardening effect that we see in our objects, then we do not expect to see 
the same effect using 2-10 keV fluxes, because they are less affected by absorption.
However, deeper observations are needed to assess this point for sources
at fluxes similar to those of the sources in our sample. The results of a 
detailed X-ray spectral analysis of the sources that were detected with 
{\it XMM-Newton} in a deep observation in the Lockman Hole field will be 
published in Mateos et al.~\cite{Mateos2005} (in preparation).

\begin{figure}
    \resizebox{\hsize}{!}{
    \includegraphics[angle=90,width=\textwidth]{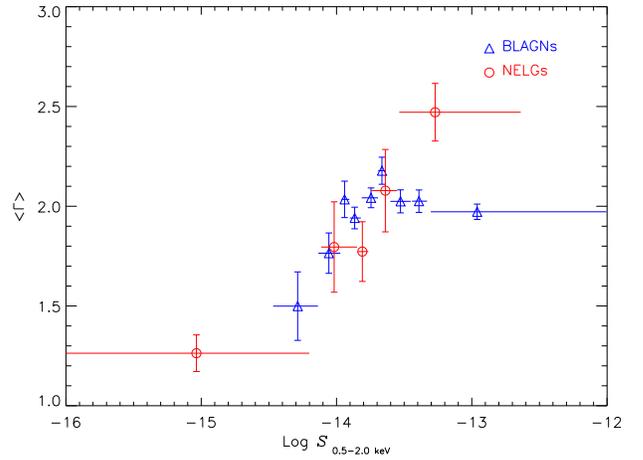}}
    \caption{$\langle\Gamma\rangle$ from best fit model versus
    $S_{0.5-2\,{\rm keV}}$ for BLAGNs (triangles) and NELGs (circles). 
    The hardening of the spectral slope is evident 
    even among the sub-samples of AGNs. Error bars correspond
    to $90\%$ confidence level.}  \label{gamma_flux_agns}
\end{figure}

The average spectral indices for the best fit model in the sub-samples of BLAGNs and NELGs 
are $1.99\pm0.02$ and
$2.35\pm0.09$, both in agreement with the canonical value measured for
nearby unabsorbed BLAGNs (Nandra \& Pounds \cite{Nandra1994}). Both
sub-samples of sources exhibit the same dependence with the soft X-ray
flux as shown in Fig.~\ref{gamma_flux_agns}. We still detect a clear
correlation between the spectral slope and the soft X-ray
flux. 

   \begin{table}[bl] \caption[]{Comparison of the weighted mean spectral photon index
   obtained for different types of sources with the values from the Maximum Likelihood 
   analysis. The best fit spectral slopes for each object were used in the calculations.} $$ \begin{array}{l c c c c c}
   \hline \noalign{\smallskip} 
   & \multicolumn{2}{c} {\rm{Maximum}} & {\rm{Weighted}}\\ 
   & \multicolumn{2}{c}{\rm{Likelihood}} &
   {\rm{Mean}}\\ {\rm Sample} & \langle\Gamma\rangle & \sigma &
   \langle\Gamma\rangle\\ \noalign{\smallskip} \hline \hline
   \noalign{\smallskip} {\rm Whole\,\,sample} & 1.86\,_{-0.03}^{+0.02} & 0.36\,_{-0.04}^{+0.01} & 1.96\pm0.01\\ 
   {\rm BLAGN} & 1.98\,_{-0.04}^{+0.04} & 0.21\,_{-0.04}^{+0.05} & 1.99\pm0.02\\ 
   {\rm NELG} & 1.85\,_{-0.17}^{+0.16} & 0.36\,_{-0.10}^{+0.14} & 2.35\pm0.09\\ 
   \noalign{\smallskip} \hline \end{array} $$
\begin{list}{}{}
\item[]
\end{list}
\label{tab3}
\end{table}

\subsection{Photon index intrinsic dispersion}
\label{intr_disp}
   \begin{figure}
    \resizebox{\hsize}{!}{
    \includegraphics[angle=-90,width=\textwidth]{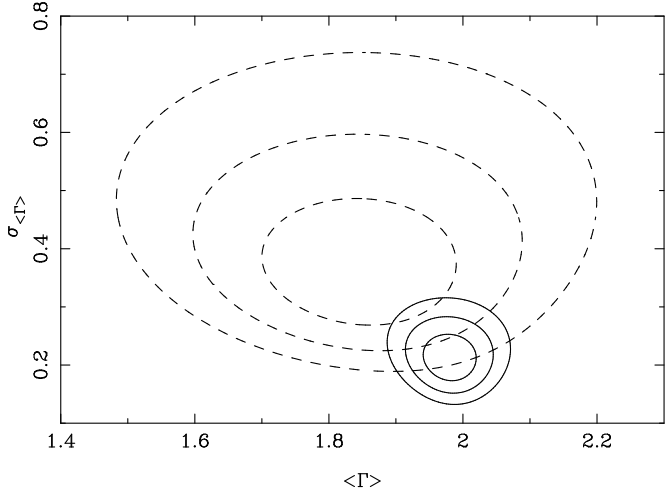}}
    \caption{Contour diagrams for the average spectral slope
    and intrinsic dispersion of our objects obtained from the Maximum 
	Likelihood analysis (see Sec.~\ref{intr_disp}).
    The spectral slopes obtained from the best fit model were used in the 
	calculations.	
    Solid line for BLAGNs and dashed line for NELGs. The contours are defined as
    $\Delta\chi^2$=2.3, 6.17 and 11.8 corresponding to standard 1,2
    and 3$\sigma$ confidence regions for two parameters.}
    \label{cont_bnef} \end{figure}

In all the above analysis we have observed a
clear dispersion in the measured spectral slope of our objects, 
even within the sub-samples of BLAGNs and NELGs, with the largest 
scatter in $\Gamma$
being observed for the NELGs. To test whether the observed scattering
is intrinsic to the sources we have followed the procedure described
in Nandra \& Pounds (\cite{Nandra1994}) and Maccacaro et al.
(\cite{Maccacaro1988}), where the distribution of source spectral
indices is assumed to be well reproduced by a Gaussian distribution of
mean $\langle\Gamma\rangle$ and dispersion $\sigma_{\langle \Gamma \rangle}$. The best
estimates of $\langle\Gamma\rangle$ and $\sigma_{\langle \Gamma \rangle}$ are obtained
with the maximum likelihood (ML) technique.  We have calculated the
distribution of the spectral slopes that we obtained from the best fit 
model for the whole sample and the sub-samples of BLAGNs and 
NELGs.  The results are shown in
Table~\ref{tab3}. In all the classes we measured a
significant dispersion in spectral slopes. The large error in 
$\Gamma$ obtained for NELGs did not allow us to differentiate the 
average spectral index of NELGs from that of BLAGNs.
We see that the mean values of $\Gamma$ computed with the ML and 
the weighted mean are very different, in particular for the NELGs. The reason 
is that with the ML method objects with untypical $\Gamma$ (i.e. outliers) 
and not very large errors will increase the dispersion of the fitted gaussian distribution 
rather than significantly affecting the mean.

Our results do not show any trend in the underlying power law of the
NELGs to be harder than that of the BLAGNs even at the faintest
fluxes. Previous results based on $ROSAT$ data (see Almaini et al
\cite{Almaini1996}, Romero-Colmenero et al. \cite{Romero1996}) 
do find a flatter
average spectral slope for NELGs ($\sim 1.5$) than for BLAGN ($\sim
2$). Our data shows that the spectral differences between these two
classes of sources are due mostly (if not totally) to differing
amounts of absorption rather than to differences in the underlying power
law spectra. Therefore many objects classified as NELGs may contain
active nuclei.

Fig.~\ref{cont_bnef} shows the $\langle\Gamma\rangle$ versus
$\sigma_{\langle \Gamma \rangle}$ contour diagrams for the sub-samples of BLAGNs (solid
lines) and NELGs (dashed lines), clearly showing the existence of 
an intrinsic dispersion in $\Gamma$ with a confidence of more 
than 3$\sigma$.

\subsection{Spectral cosmic evolution}

\begin{figure}
    \hbox{
    \includegraphics[angle=90,width=8.5cm]{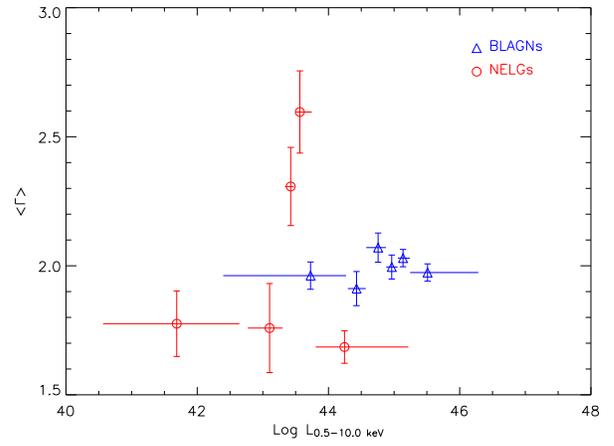}}
    \caption{Weighted $\langle\Gamma\rangle$ versus ${\rm
    L}_{0.5-10\,{\rm keV}}$ for BLAGNs (triangles) and NELGs
    (circles). Error bars correspond to $90\%$ confidence level.}
    \label{gamma_var}
\end{figure}

\label{cosmic_evolution}
\begin{figure}
    \resizebox{\hsize}{!}{
    \includegraphics[angle=90,width=\textwidth]{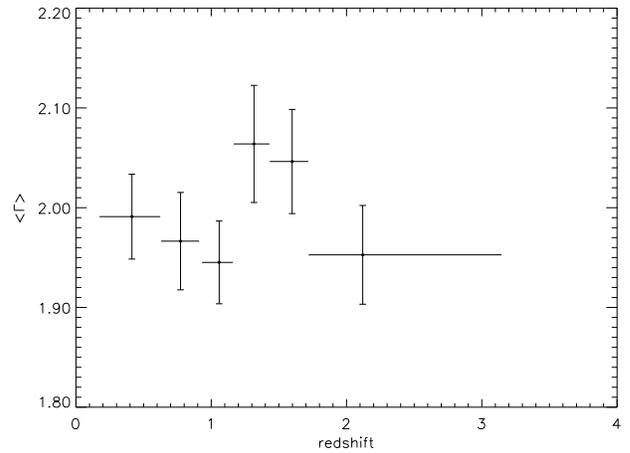}}
    \caption{$\langle\Gamma\rangle$ versus redshift for BLAGNs. We have used the values of 
      $\Gamma$ from the best fit model for each object (i.e. with absorption and soft excess 
      if required). Error bars correspond to $90\%$
    confidence level.}  \label{gamma_z_2}
\end{figure}

No correlation was found between $\Gamma$ and the 
rest-frame 0.5-10 keV luminosities, as shown in
Fig.~\ref{gamma_var}.  For the sub-sample of BLAGNs we studied the
measured $\Gamma$ that we obtained from each best fit model as a function 
of the redshift, but we did not find any clear tendency between the two
parameters as can be seen in Fig.~\ref{gamma_z_2}.

\begin{figure}
    \resizebox{\hsize}{!}{
    \includegraphics[angle=90,width=\textwidth]{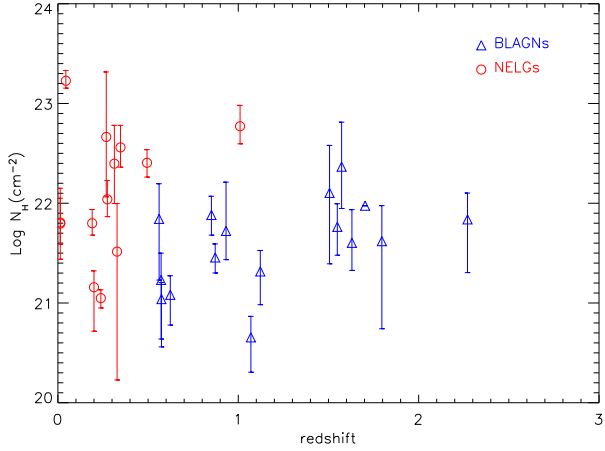}}
    \caption{Intrinsic absorption as a function of redshift for 
      absorbed (F-test significance $\ge$95\%) BLAGNs (triangles)
    and NELGs (circles).}  \label{nh_z}
\end{figure}

We have studied the redshift distribution of the measured rest frame
absorbing column densities for the absorbed sub-samples of BLAGNs and
NELGs.  The results, shown in Fig.~\ref{nh_z}, show 
no evidence of 
evolution of ${\rm N_H^{intr}}$ with redshift. However, the 
ranges of the 2-D parameter space that we fully sample with our source list 
are unclear.

We expect the minimum detected absorption to increase with $z$. This  
can be easily explained in terms of absorption being redshifted towards
lower energies and further decreased by a factor $(1+z)^{2.7}$ as we move
upwards in $z$. 

To study if there exists a maximum detected column (only the brightest 
sources being actually detectable when heavy absorption is present), and if 
this limit depends on redshift, we have carried out simulations as follows. 
A 3-D parameter space was defined with the axes being redshift,
luminosity and intrinsic (rest-frame) absorption. For each parameter
we have defined a grid of values covering the observational limits
established by our sources.  First, for each point in the ${\rm
N_H^{intr}}$-$z$ plane we simulate a spectrum for each BLAGN or NELG
source in our sample, using its spectral response matrices and
exposure time and assuming $\Gamma=2$.  As explained in
Sec.~\ref{products} we have only analysed those spectra with
more than 50 counts (background subtracted). We use this limit to
calculate the required minimum luminosity for each source to fulfil
this constraint. Luminosities lower than this one would produce a
spectrum outside our quality bounds.

Second, for a number of luminosities above this limit (still, for each
value of intrinsic absorbing column and $z$) we simulate a spectrum
for every source. These spectra were analysed following the same
procedure that we used for the real spectra giving 
a full coverage of the above 3-D parameter space.

We then examined the fitted values of ${\rm N_H^{intr}}$ as a function
of redshift $z$ for a range of luminosities. The simulations confirm
that the minimum detection of absorption depends strongly on $z$ for
the reason explained above. We also found two limits to the maximum
absorption detected as a function $z$. The first one is the source
detection itself: heavy absorption reduces significantly the flux of
the source and makes detection very difficult, especially at
high redshifts. The second limit is the detection of absorption: even
if the source is detected, if its luminosity is not high enough we
will measure an absorption below the maximum value
established by the detection limit.

This strong dependence on luminosity and redshift makes it 
very difficult to derive the real absorbing column distribution in
terms of the fitted values.  The best way we found to tackle this
problem is to compare our results with the predictions of specific
models for the distribution of absorbing columns, folding the input
values through the same effects as the real data. This is discussed in
Sect.~\ref{synthesis} in the framework of unified AGN models for the
XRB.

\section{ALGs X-ray spectra}
\label{ALGs}

The objects with galaxy-like optical spectra not showing emission line 
signatures were classified as
absorption line galaxies (ALG).

The inferred high X-ray luminosities obtained from the analysis of their 
X-ray spectra strongly suggests that an AGN is the source of the X-ray emission 
in at least 5 out of the 7 objects. The peculiar properties of these objects 
have been explained by a very high column density that would be obscuring the AGN broad
and narrow line regions. This hypothesis has found observational support, 
because many of the ALGs studied have been found to be 
X-ray obscured (Mainieri et al. \cite{Mainieri2002}, Comastri et al. 
\cite{Comastri2002}). However there are several examples of ALGs for 
which the obscuration model does not apply. In these cases 
other possible explanations have been suggested, e.g. that a BL Lac could be 
present in the core of some ALGs (Brusa et al. \cite{Brusa2003}). Alternatively some objects might  
host an AGN with a non-standard accretion disk (Yuan \& Narayan \cite{Yuan2004}).  
A final possibility is that some ALGs may have normal (unabsorbed) AGNs that could be hidden 
if the host galaxy is bright enough (Severgnini et al. \cite{Severgnini2003}, Page et al. \cite{Page2003}).

We now describe
the X-ray spectral properties of the ALGs found in our list:

{\bf XMMU J122017.7+752216} is an optically normal galaxy that looks  
extended in the X-ray image. Its soft X-ray
emission is best reproduced with a thermal model with temperature
$k\rm{T=0.64_{-0.04}^{+0.04}}$ keV and metallicity $Z=Z\odot$. The fit
does not improve leaving $Z$ as a free parameter. However, the 
high energy X-ray spectrum is best modelled with an
unabsorbed power law of spectral slope $\Gamma=1.87_{-0.11}^{+0.12}$. We cannot 
confirm whether the source of the hard X-ray emission is an AGN
due to the low value of the hard X-ray luminosity inferred from this
model ($\rm{L_{2-10}=3.1\times10^{39} erg\,s^{-1}}$).

The nature of {\bf XMMU J021822.1-050612} was studied in detail
by Severgnini et al. (\cite{Severgnini2003}). They found the X-ray
spectrum to be best modelled with an absorbed power law.

The X-ray spectra of {\bf XMMU J010320.2-064157}
and {\bf XMMU J084211.6+710145} were best
reproduced with an obscured (F-test values 99.7\% and 95.0\%) AGN with
$\Gamma=2.39_{-0.96}^{+1.36}$, ${\rm
N_H=4.9_{-2.53}^{+5.28}\times10^{22}\,cm^{-2}}$ and ${\rm
L_{2-10}=2\times10^{44}\,erg\,s^{-1}}$ for the first object, and
$\Gamma=1.82_{-0.15}^{+0.16}$, ${\rm
N_H=9.3_{-6.8}^{+8.5}\times10^{20}\,cm^{-2}}$ and ${\rm
L_{2-10}=3.88\times10^{43}\,erg\,s^{-1}}$ for the second object.

{\bf XMMU J214041.3-234719} has an unabsorbed X-ray spectrum
with $\Gamma=2.08_{-0.08}^{+0.09}$ and ${\rm
L_{2-10}=6.2\times10^{40}\,erg\,s^{-1}}$. 
Hence, from its X-ray spectrum alone, we would expect this source 
to be a BLAGN optically, rather that an ALG, according to the simplest AGN unified model.

\begin{figure}
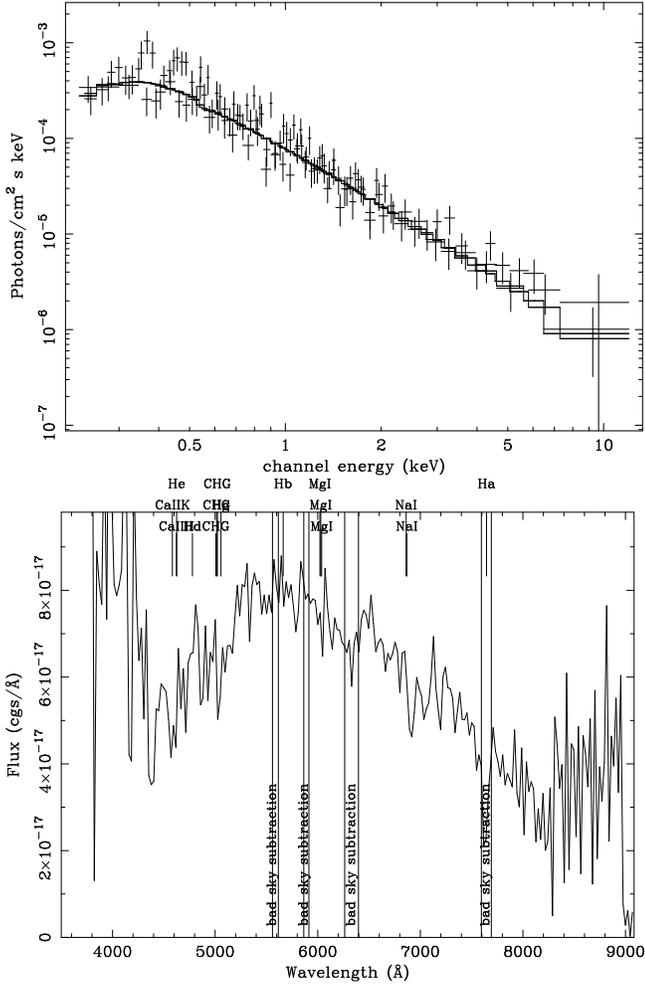

	 \hbox{
   \includegraphics[angle=-90,width=8.5cm]{fig25.ps}}
	\hbox{
   \includegraphics[angle=-90,width=8.5cm]{fig26.ps}}
   \caption{Top panel: MOS+pn spectrum of the absorption line galaxy {\bf XMMU
   J233113.6+20056} at z=0.378. The X-ray spectrum was best fitted with a flat ($\Gamma\sim0.944$) 
   unabsorbed power law. Bottom panel: optical spectrum of the source, where no obvious broad lines were detected} 
   \label{alg_spec} \end{figure}

A simple power law model can reproduce the spectra of {\bf
XMMU J233113.6+20056} (see Fig.~\ref{alg_spec}) 
and {\bf XMMU J122312.6+75247}, but the best fit
spectral slopes were found to be well below the typical value found
for unabsorbed AGNs ($\Gamma=0.94_{-0.27}^{+0.26}$ and
$\Gamma=1.36_{-0.28}^{+0.30}$). In terms of the F-test none of the
spectra required absorption in excess of the Galactic value, although
with the absorbed model we obtained more typical spectral slopes.  
The low quality of the spectra has not allowed us to
study in more detail whether the spectra of these sources is
intrinsically flat or absorbed.

\section{Testing population synthesis models}
\label{synthesis}

The standard population synthesis models of the XRB, based on
unification schemes for AGN, can reproduce successfully the XRB with
an appropriate mixture of unobscured and obscured AGN, as well as a
number of observational constraints. The input parameters of these
models are the type 1 and type 2 measured X-ray spectra and
cosmological evolution, which is assumed to be identical.  The X-ray
luminosity function of absorbed objects is still unknown, but it is
assumed to be the same as for unabsorbed objects.  The ratio of
absorbed to unabsorbed AGN comes out to be rather large in these
models.  A key descriptor of these models is the distribution of
absorbing column densities among the sources. The ${\rm N_H^{intr}}$
distribution is now observationally known for the local Seyfert
galaxies (e.g., Risaliti et al. \cite{Risaliti1999}), and in general
these models assume it to be independent of redshift and luminosity.

The analysis that we have conducted has allowed us to measure the
distribution of absorption in one of the largest samples of BLAGNs and
NELGs studied up to now, and its cosmic evolution over a wide range of
redshifts (see Fig.~\ref{nh_in} and Fig.~\ref{nh_z}).  In
Sec.~\ref{cosmic_evolution} we reported that the measured absorption is
slightly lower than the real one, depending on the luminosity and
redshift of the source. Hence, in order to compare the predictions of
XRB models to our data, we assume an input distribution for ${\rm
N_H^{intr}}$ (independent of luminosity and redshift), and then fold it 
through the same detection effects as in our data.

We have used the ${\rm N_H^{intr}}$ distributions adopted by Comastri
et al. \cite{Comastri1995}, Gilli et al. \cite{Gilli2001} and 
Pompilio, La Franca \& Matt \cite{Pompilio2000}, as input parameters to their analytical
models of the XRB.  We have simulated spectra for each one of the
BLAGNs and NELGs included in our list, by fixing their redshifts,
exposure times, position and luminosities.  The slope of the
continuum was assumed to be $\Gamma=2$ for all the sources.  For each
source we extracted ${\rm N_H^{intr}}$ values from the specific model
distribution, until the simulated spectrum fulfilled the quality
filters that we used for the real spectra (binned spectra with at
least 5 bins and source net counts above 50). Indeed, heavily obscured
sources do not pass, just as happens for real sources. We fitted the
spectra with model A and model C and computed F-test significances for
absorption, using the same calibration matrices.

\begin{figure*}
   \hbox{
   \includegraphics[angle=90,width=8.5cm]{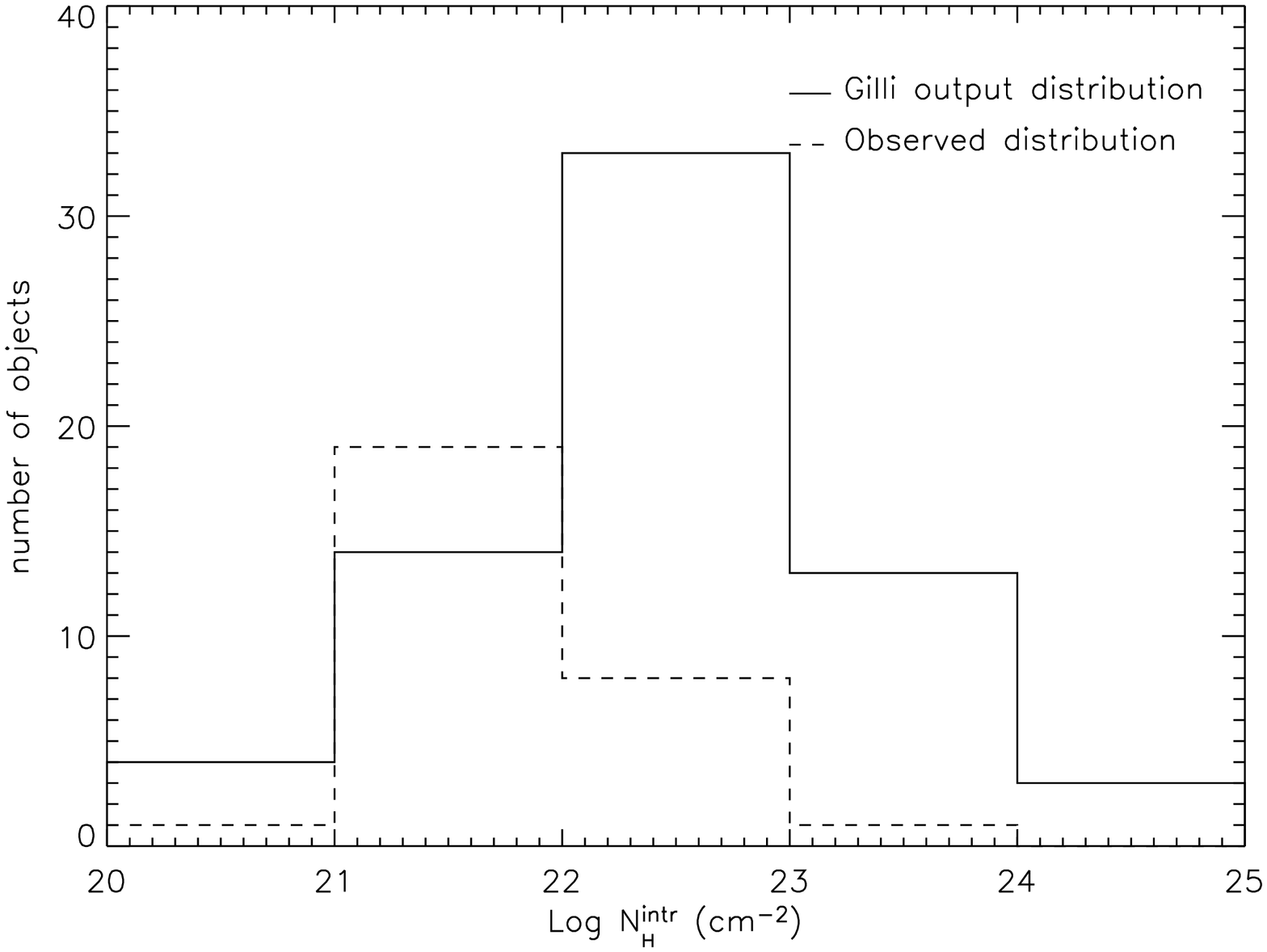}
   \includegraphics[angle=90,width=8.5cm]{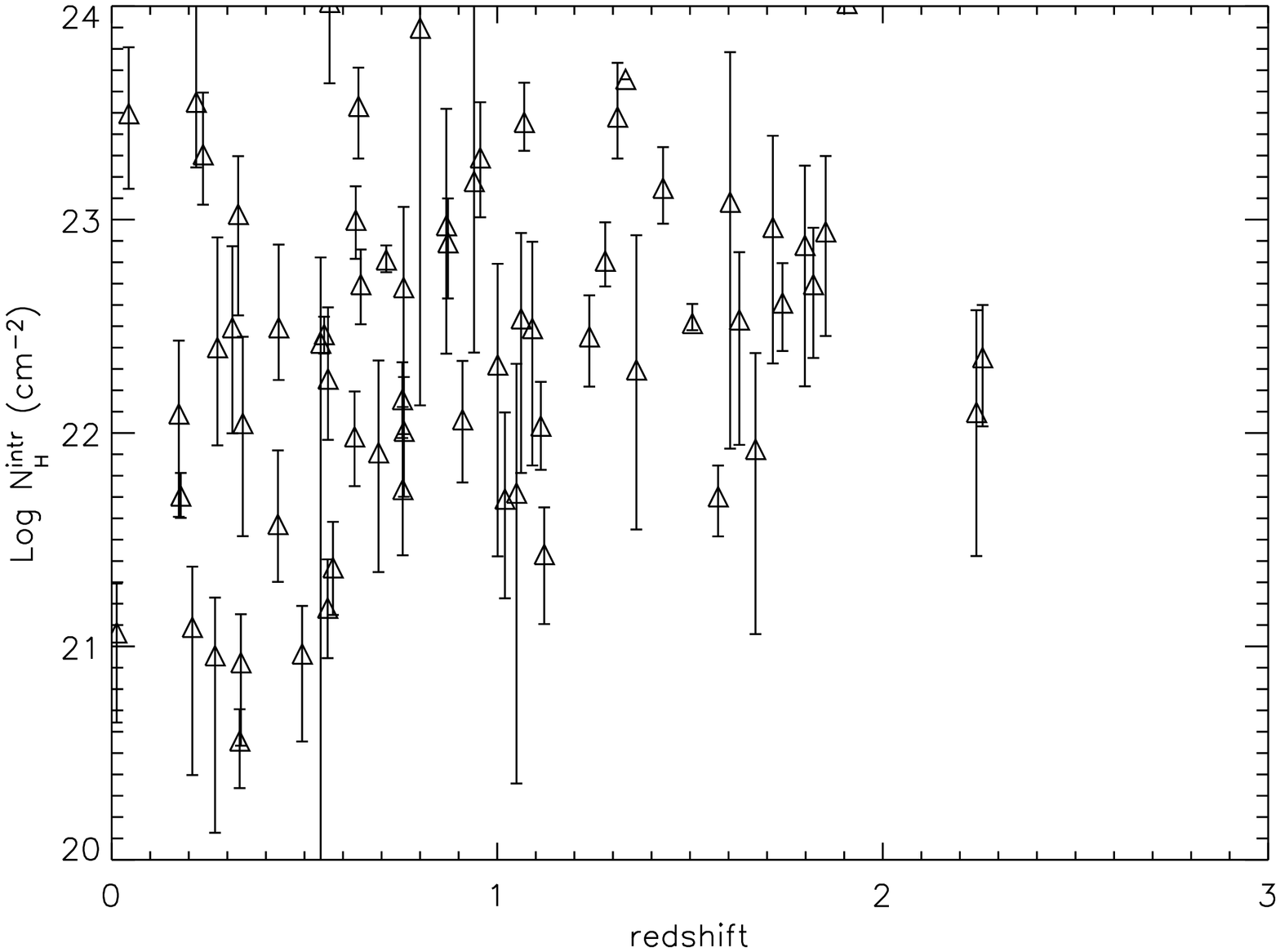}}
   \caption{Left: Histograms of column density distributions in absorbed objects  
   obtained from a simulated input distribution of ${\rm N_H^{intr}}$
   according to the model of Gilli et al. \cite{Gilli2001}.  The dashed
   histogram represents the observed distribution for our sources.
   Right: ${\rm N_H^{intr}}$ as a function of sources' redshifts from simulations.
   }  \label{sims}
\end{figure*}

Fig~\ref{sims} compares the values of ${\rm N_H^{intr}}$ measured in
the real data and in the simulations for one of the XRB models (Gilli
et al. \cite{Gilli2001}). Obviously, the adopted distribution
is not able to reproduce the observational results. Very similar
graphs are obtained for the models of Comastri et al
\cite{Comastri1995} and Pompilio et al. \cite{Pompilio2000}.  For all
three models there is a clear excess of sources with intrinsic
absorptions ${\rm N_H^{intr}}\ge 10^{22}\,{\rm cm}^{-2}$, which are not
present in the real data.  For an F-test confidence level of $\ge95\%$
we detect, in our real sample, absorption in 29 sources out of 179.  The number of absorbed
sources detected in the simulated data sets were 79, 67 and 71 for the
Comastri et al. \cite{Comastri1995}, Gilli et al. \cite{Gilli2001} and
Pompilio et al. \cite{Pompilio2000} distributions, in all cases a
factor of more than 2.5 over the observed value. This result is also
evident from Fig~\ref{sims} (right panel), where the predicted distribution of absorbing 
columns with the Gilli et al. \cite{Gilli2001} model is plotted against redshift. We find 
excess of absorbed objects all
over the redshift interval up to redshifts of $\sim$2. 
Discrepancies between the predictions from synthesis models 
and what it is obtained from observations have been observed previously (i.e. 
Piconcelli et al. \cite{Piconcelli2002}, 
Caccianiga et al. \cite{Caccianiga2004}, Georgantopoulos et al. \cite{Georgantopoulos2004}, 
Perola et al. \cite{Perola2004}). 

In the recent ``modified 
unified scheme'' of Ueda et al. \cite{Ueda2003} the AGN absorbing column density distribution 
depends on luminosity (not redshift dependence).  
The fraction of absorbed AGNs (${\rm N_H>10^{22}\,cm^{-2}}$) decreases significantly with luminosity. 
This model will reduce the observed discrepancies
fundamentally at high luminosities, and therefore at high redshifts. However, this model predicts 
a significant fraction of objects with ${\rm N_H>10^{22}\,cm^{-2}}$ at the typical luminosities 
of our sample that we do not detect.  The Ueda et al. \cite{Ueda2003} model, although more in 
line with our results, still overpredicts the number of absorbed objects (${\rm N_H>10^{22}\,cm^{-2}}$) especially 
 at low redshifts.

One possible caveat is that some of these absorbed sources that we do
not detect in the real spectra might be hiding within the
unidentified sources. Our simulations indicate that $\sim 40\%$ should
have detectable absorption. However, even for the full sample of 1137 
objects we only detect absorption in 10-20\% of them. Assuming that the
redshift distribution of the identified sources does not strongly
deviate from the true one (and there is evidence that this is the case
from other surveys), then 40\% of absorbed sources predicted by
the simulations should also be found in the whole sample, while only
$\sim 20\%$ are seen.  This confirms that XRB synthesis models predict
twice as many absorbed sources at our flux levels than we actually
observe.

\section{Conclusions}
\label{conclusions}

We have conducted a detailed analysis of the broad band spectral
properties of a large sample of sources detected serendipitously with
the {\it XMM-Newton} observatory.  Various spectral models have been tested
to reproduce the 0.2-12 keV emission of our sources. We summarise
our main conclusions as follows:

\begin{enumerate}
   \item Fitting the spectra of all the sources with a single power
   law we obtain a weighted mean of $\langle\Gamma\rangle=1.86\pm0.02$, 
   in agreement
   with previous surveys. However, for a significant number of sources, the
   quality of their fits was very poor, with 7-12\% of the sources having 
   statistically unacceptable fits (null hypothesis probability
   $< 5\%$). 
   \item We observed the continuum average spectral
   slope to become harder with decreasing 0.5-2 keV flux. This effect
   does not appear to be correlated with the quality of the spectral 
   data. After correcting for all the biases that could be affecting 
   our results, we still see the same dependence, 
   therefore the hardening of $\Gamma$ with decreasing soft flux is real.

   \item Many objects
   in our sample exhibit absorption in excess of the Galactic value. In terms of 
   the F-test (95\% confidence level), excess absorption is found in 
   17\% of sources. Significant fractions of X-ray
   absorbed objects were found among NELGs ($\sim$ 40\%), BLAGNs
   ($\sim$ 6\%) and ALGs ($\sim$ 40\%). This result clearly suggests that the relationship between
   optical identification and X-ray spectral signatures is still
   unclear and needs some revision. 

\item The distributions of absorbing columns in BLAGNs and NELGs do not
appear significantly different within the limitations of our sample, except for the overall factor that
NELGs are 4 times more frequently absorbed than BLAGNs. 

\item The typical absorption
   columns derived for BLAGNs (${\rm N_H< 10^{22}\, {\rm cm}^{-2}}$) indicate
   that perhaps the host galaxy, or some sort of weak distant absorber
   might provide the absorbing gas, without blocking too much the line
   of sight to the Broad-Line region. This point deserves further
   exploration.

\item Using the best fit model for each object, i.e., including absorption 
   and soft excess 
   when required by the data, we still observe a substantial hardening of the 
   average spectral index towards fainter soft fluxes. The same effect
   was also observed for the sub-samples of BLAGNs and NELGs. With the
   current data quality, we cannot reject that these sources have the
   same underlying power law ($\Gamma\sim 2$) with moderate absorption,
   nor can we reject that there is a population with genuinely flatter
   spectral index. However when hard fluxes are used to study this effect 
   there is no evidence of a change of average spectral slope. This
   supports the hypothesis that absorption is producing the hardening of the 
   average spectra of our sources, because hard band fluxes are less affected by absorption.

\item Using a maximum likelihood fit we have detected an intrinsic dispersion of the spectral slope, even 
   within the sub-samples of BLAGNs and NELGs, with the largest scattering observed for the 
   NELGs.
 
  \item We did not find a tendency of absorbed BLAGNs to occur preferentially at any particular redshifts. The same 
   result is obtained when comparing the 2-10 keV luminosity distributions of absorbed and unabsorbed 
   BLAGNs and NELGs. Therefore, no evidence is found of unabsorbed BLAGNs and 
   NELGs having different X-ray luminosities to the absorbed objects. 

\item We have seen 
   that our unabsorbed NELGs are on average less luminous than the unabsorbed BLAGNs and hence 
   our data are consistent with the explanation of the lack of broad lines in the optical spectra 
   of unabsorbed NELGs as an AGN/ host-galaxy contrast effect.

\item A soft excess component was detected with a high significance
   in $\sim 7\%$ of BLAGNs and NELGs. 

\item For XRB synthesis models based on the AGN unified
scheme, simulations were conducted to test the distribution of
absorbing columns from these models on the population of BLAGNs and
NELGs. In general, the number of sources where absorption was detected
in the simulated spectra (in about $\sim 40\%$ of them) was a factor
of $\sim2.5$ above the measured values for all models.  Regardless of
the incompleteness of the identifications, the XRB synthesis model
prediction of detectable absorption in 40\% of the sources is at odds
with the overall observed rate of absorbed sources in the
real sample.

\item The distribution of absorbing column densities predicted by the
synthesis models and measured through the simulations is significantly
skewed towards high column densities ${\rm N_H^{intr}\sim
10^{22}-10^{24}\,cm^{-2}}$. These highly absorbed sources, predicted
by the XRB models at our flux levels, are not detected in our sample of 
objects.

\end{enumerate}

\begin{acknowledgements}
 SM acknowledges support from a Universidad de Cantabria fellowship. SM,
 XB, FJC and MTC acknowledge financial support from the Spanish Ministerio
 de Educaci\'on y Ciencia, under projects AYA2000-1690 and ESP2003-00812.  
This
 project was supported in part by the German BMBF under DLR grant 50
 OX 0201. AC and TM acknowledge financial support from Italian Ministry
of University and Scientific and Technological Research (MURST) through 
grant Cofin. We also thank the anonymous referee for his/her suggestions that helped 
to improve the manuscript considerably.  

\end{acknowledgements}

\appendix
\section{Sources' detection efficiency function}
\label{apendix_A}

The use of different energy bands with different sensitivities to detect our sources, the selection 
of fields covering a wide range of exposure times (from $\sim$ 10 to $\sim$ 100 ksec),
and the quality filters applied to the spectra (especially the selection of spectra with 
MOS+pn counts $\ge$ 50) makes the calculation of the efficiency of source detection as 
a function of flux difficult. We know it will be a function of the X-ray flux, 
but it is important to know 
whether it also depends on the spectral shape of the sources.

We have conducted simulations to calculate the efficiency of detection, 
W($\Gamma$, $\it S$): for a given number of sources with spectral parameters $\Gamma$ and 
$\it S$, W($\Gamma$, $\it S$) gives us the fraction of simulated objects that were detected, i.e., 
their spectra fulfilled the quality filters that we applied to our data, see Sec.~\ref{products}).
and with best fit spectral parameters being the same as the input (simulated) ones.
This function can be used to correct observed source properties, for example the 
dependence of $\langle \Gamma \rangle$ with the X-ray flux, from all the 
effects listed above.
We defined a grid of points in $\Gamma$-$\it S$ covering the range of 
values measured for our objects ($\Gamma$ from 0.5 to 3; {\it S} from ${\rm 10^{-16}}$ to 
${\rm 10^{-12}\,erg\,cm^{-2}\,s^{-1}}$).

We have simulated the same 
number of spectra on each grid point. 
For these simulations we have used the list of detected 
sources before applying the quality filters, i.e., 2145 sources have been simulated for 
each point in the $\Gamma$-{\it S} grid.
The input parameters of each simulated spectrum were the exposure time, 
response matrices and Galactic 
absorption of each real source, and the values of $\Gamma$ and {\it S} at each grid point. 
We used model A (see Sec.~\ref{modA}) to simulate all the spectra,  
therefore absorption is not included in the simulations.
Once all the spectra were simulated, we applied the same quality filters that we used for our real 
data. Our source detection efficiency, at each grid point, was defined as 
\[
W(\Gamma, {\it S})={N_{det} (\Gamma, {\it S}) +\, N_{det} (\Gamma', {\it S}') \over N_{sim}}
\]
where $N_{sim}$ is the number of sources that were simulated at the grid point ($\Gamma$, {\it S}) (this 
number is the same for all the grid points). It is important to notice that, there are two possible 
contributions to the final number of sources that will appear at a given grid point ($\Gamma$, {\it S}):
\begin {enumerate}
\item The number of simulated sources at the grid point ($\Gamma$, {\it S}) that, after fitting, 
remain at 
      the same grid point, $N_{det} (\Gamma, {\it S}$).
\item The number of sources simulated at different grid points ($\Gamma$', {\it S}') that were 
      detected, and with best fit spectral parameters ($\Gamma$, {\it S}), $N_{det} (\Gamma', {\it S}')$.
\end {enumerate}

\begin{figure}
    \hbox{
    \includegraphics[angle=90,width=8.5cm]{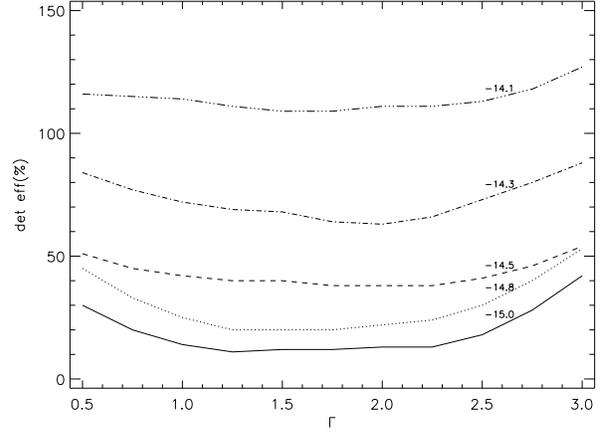}}
    \caption{ Dependence of the detection efficiency of sources as a function of $\Gamma$ at 
	different flux levels, obtained from simulations. Some values of the efficiency function can 
	be above 100\% because the fitted parameters may differ from the input ones (see text for details). 
	}
    \label{selection_funct}
\end{figure}

\begin{figure}
    \hbox{
    \includegraphics[angle=90,width=8.5cm]{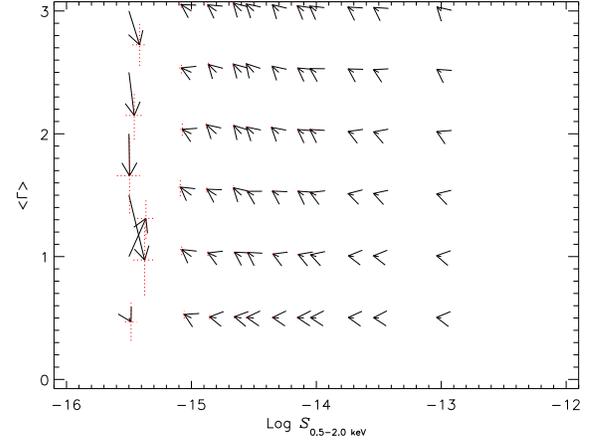}}
    \caption{Representation of the average movements in the $\Gamma$-{\it S} plane obtained from 
	simulations. The beginning point of each arrow indicates the input values 
	of $\Gamma$ and {\it S} in the simulations; the end points are the observed (fitted) 
	average values of $\Gamma$ and {\it S} for the detected sources.  
	}
    \label{arrows}
\end{figure}

The values of the detection efficiency as a function of $\Gamma$ are plotted in 
Fig.~\ref{selection_funct} for different fluxes. We observe that, at fluxes below 
$\rm {\sim 10^{-14}\,erg\,cm^{-2}\,s^{-1}}$, the efficiency of detection 
decreases significantly from $\Gamma$=0.5 to $\Gamma$=2, and then, starts to increase up 
to $\Gamma$=3. The rise of efficiency at high values 
of $\Gamma$ could be due to the sharp increase in the effective areas of the EPIC detectors at low energies 
combined with the increase in the relative contribution to the total counts at these energies for 
the steeper spectral slopes.
It is interesting to notice that, although we have shown that the detection efficiency increases for $\Gamma$ values above 2.5, 
we have not detected these super-soft sources in our sample, hence, they must be rare. The reason why 
the efficiency function is higher for objects with hard ($\Gamma\sim$0.5) spectral slopes 
(notice that the effective area of the X-ray detectors drops rapidly at energies below $\sim$1 keV) is 
that a source with a flat spectral slope has to be much brighter than one with a soft 
($\Gamma\sim$2) spectral slope in order to have the same 0.5-2 keV flux. Therefore we will receive 
more counts in the 0.2-12 keV band from flatter objects, which means that it will better detect them
and more easily pass our spectral quality filters.

Our simulations showed that there are movements of objects in the  $\Gamma$-{\it S} plane, i.e., 
the fitted parameters may differ from the input ones.
It is necessary to study how important this effect is and whether there are systematic shifts 
(for example a tendency to detect the sources with lower values of gamma at faint fluxes). Hence, we have calculated for 
each grid point the average values of $\Gamma$ and $\it S$, in the same way as we did for the real sample,
 from all the sources simulated 
at that point that were detected (no matter at which grid point they were detected). The results, shown in
Fig.~\ref{arrows}, confirm the existence of shifts in the $\Gamma$-{\it S} plane at the faintest fluxes: on average, the 
measured $\Gamma$ values tend to be softer, and hence the observed 0.5-2 keV fluxes are lower. However,  
for fluxes $\rm {\ge 10^{-15}\,erg\,cm^{-2}\,s^{-1}}$ these movements are not significant and tend 
to soften than harden spectra.

\end{document}